\documentclass[12pt]{article}

\pdfoutput=1

\usepackage{color}
\usepackage{amsmath}
\usepackage{amsfonts}
\usepackage{amssymb}
\usepackage{caption}
\usepackage{graphicx}
\usepackage{slashed}            
\usepackage{subfig}             
\usepackage{xspace}				
\usepackage{cite}

\usepackage[english]{babel}
\usepackage{fancyhdr}
\usepackage{amsmath}
\usepackage{amssymb}
\usepackage{amsfonts}
\usepackage{psfrag}
\usepackage[applemac]{inputenc}
\usepackage{graphicx}

\def\gsim{\mathrel{\rlap{\lower4pt\hbox{\hskip1pt$\sim$}}
    \raise1pt\hbox{$>$}}}                

\newcommand{\arXiv}[2]{\href{http://arxiv.org/pdf/#1}{{\tt #2/#1}}}
\newcommand{\arXivold}[1]{\href{http://arxiv.org/pdf/#1}{{\tt #1}}}

\newcommand{\GeV}{\,\mathrm{GeV}}

\newcommand{\beq}{\begin{eqnarray}}
\newcommand{\eeq}{\end{eqnarray}}

\newcommand{\bag}{\begin{align}}
\newcommand{\eag}{\end{align}}

\newcommand{\MeV}{\,\mathrm{MeV}}

\newcommand{\vev}[1]{\langle {#1} \rangle}

\usepackage[hypertexnames=false]{hyperref}		

\begin{document}
\begin{titlepage}

\vskip.5cm

\begin{center} 
{\huge \bf Cosmological and Astrophysical Probes  of Vacuum Energy \vspace*{0.3cm}   } 
\end{center}

\begin{center} 
{\bf \  Brando Bellazzini$^{a,b}$, Csaba Cs\'aki$^c$, Jay Hubisz$^d$, \\ Javi Serra$^b$, and John Terning$^e$} 
\end{center}

\begin{center} 

$^a$ {\it Institut de Physique Th\'eorique, CEA Saclay,
 F-91191 Gif-sur-Yvette, France}\\

\vspace*{0.1cm}

$^b$ {\it Dipartimento di Fisica e Astronomia, Universit\`a di Padova \& INFN, Sezione di Padova, I-35131 Padova, Italy}\\

 \vspace*{0.1cm}

$^{c}$ {\it Department of Physics, LEPP, Cornell University, Ithaca, NY 14853, USA} \\

\vspace*{0.1cm}

$^{d}$ {\it Department of Physics, Syracuse University, Syracuse, NY  13244} \\

\vspace*{0.1cm}

$^{e}$ {\it Department of Physics, University of California, Davis, CA 95616} \\

\vspace*{0.1cm}

{\tt  
 \href{mailto:brando.bellazzini@pd.infn.it}{brando.bellazzini@cea.fr},
\href{mailto:csaki@cornell.edu}{csaki@cornell.edu}, \\
 \href{mailto:jhubisz@phy.syr.edu}{jhubisz@phy.syr.edu},  
 \href{mailto:jserra@pd.infn.it}{jserra@pd.infn.it},
 \href{mailto:jterning@gmail.com}{jterning@gmail.com}}

\end{center}

\vglue 0.3truecm

\centerline{\large\bf Abstract}
\begin{quote}
Vacuum energy changes during cosmological phase transitions and becomes relatively important at epochs 
just before phase transitions. For a viable cosmology the vacuum energy just after a phase transition must be set by the critical temperature of the next phase transition, which exposes the cosmological constant problem from a different angle. Here we propose to experimentally test the properties of vacuum energy under circumstances different from our current vacuum. One promising avenue is to consider the effect of high density
phases of QCD in neutron stars. Such phases have different vacuum expectation values and
a different vacuum energy from the normal phase, which can contribute an order one fraction to the mass of neutron stars.  Precise observations of the mass of neutron stars can potentially yield information about the gravitational properties of vacuum energy, which can significantly affect their mass-radius relation.  A more direct test of cosmic evolution of vacuum energy could be inferred from a precise observation of the primordial gravitational wave spectrum at frequencies corresponding to phase transitions. While traditional cosmology predicts steps in the spectrum determined by the number of degrees of freedom both for the QCD and electroweak phase transitions, an adjustment mechanism for vacuum energy could significantly change this. In addition, there might be other phase transitions where the effect of vacuum energy could show up as a peak in the spectrum. 

 \end{quote}

\end{titlepage}

\tableofcontents

\setcounter{equation}{0}
\setcounter{footnote}{0}

\section{Introduction: A brief history of vacuum energy}
\setcounter{equation}{0}

The discovery of the acceleration of the Universe~\cite{accelerating} has led to one of the deepest puzzles of modern day physics. Within cosmology the dark energy responsible for the acceleration can simply be described by adding a new parameter, the cosmological constant, to the expansion equations.  However, within particle physics this cosmological constant is expected to correspond to the vacuum energy of the quantum field theory of our Universe, determined by the underlying microscopic physics. It is then difficult to explain why a simple estimate for the vacuum energy is many orders of magnitude larger than the observed value, $\Lambda \sim (10^{-3} \ {\rm eV})^4$, which is much smaller than any other scales appearing in the Standard Model (SM) of particle physics. Supersymmetry (SUSY) is the only known mechanism to set the cosmological constant to zero, however SUSY breaking does contribute to the vacuum energy, resulting in the oft quoted 60 orders of magnitude discrepancy known as the cosmological constant problem. On the other hand, if there is a (yet to be identified) adjustment mechanism for the cosmological constant,\footnote{Any such adjustment mechanism is strongly constrained by the Weinberg no-go theorem~\cite{Weinbergnogo}, for recent discussions see~\cite{dilatoncc}.} then why is it not exactly zero? This has led many scientists to embrace Weinberg's approach, who predicted the expected magnitude of the cosmological constant from anthropic considerations: if the cosmological constant was much larger than the critical density then structure could not have formed, given the observed size of primordial density perturbations.

\begin{figure}[!t]
\begin{center}
\includegraphics[width=3.1in]{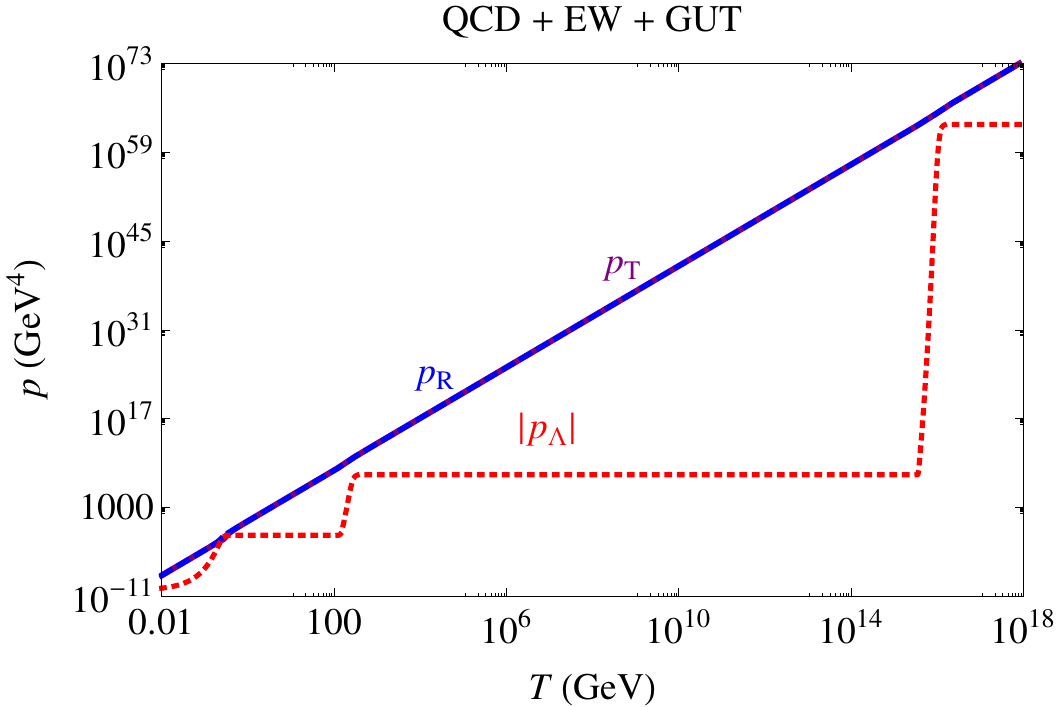} \includegraphics[width=3.1in]{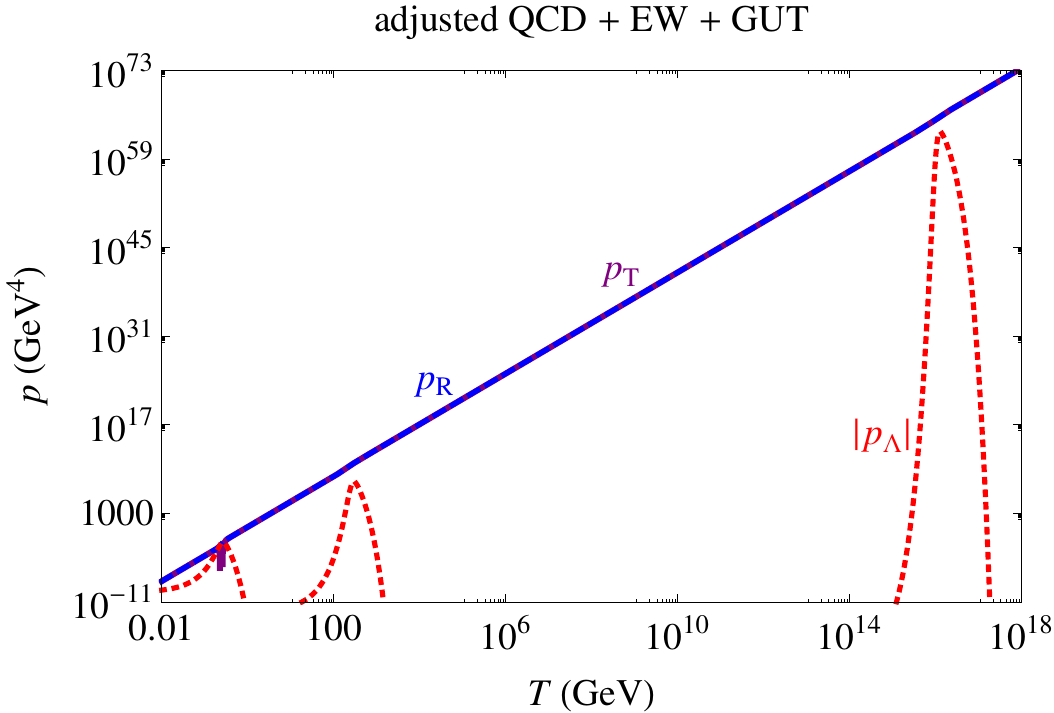}
\caption{Sketch of the evolution of vacuum energy (dotted-red) and the total pressure (solid-purple) dominated by radiation (dashed-blue) during the expansion of the Universe. Left: standard model evolution where the vacuum energy jumps at every PT (the ones pictured here correspond to the GUT, EW and QCD PTs). Right: the evolution assuming some form of adjustment mechanism for vacuum energy. \label{fig:evolution}}
\end{center}
\end{figure}

Looking at the cosmic history of the Universe, one can realize that the cosmological constant problem is actually more severe than the tuning of a single parameter.  At every phase transition (PT) the Universe undergoes (when the vacuum expectation values of fields are changing), the vacuum energy is expected to jump by an amount proportional to the critical temperature $T_c$ \cite{Bludman:1977zz,Weinbergnogo}:
\begin{equation}
\Delta \Lambda_i \propto T_{c,i}^4 \ .
\end{equation} 
In order for vacuum energy to not dominate after the PT (and thus allow ordinary radiation dominated expansion of the Universe in accordance with successful structure formation), the total vacuum energy after the end of the PT has to be quite precisely equal to the change in vacuum energy generated at the {\it next} PT. Viewed from this angle the cosmological constant problem is even more disturbing: every time vacuum energy is about to dominate the energy density, a new PT must happen, and the amount of cancellation of vacuum energy during the PT already anticipates the future history of the Universe. For example at temperatures above the electroweak (EW) scale the vacuum energy in the SM is of order $M_W^4$. As the Universe cools and goes through the EW PT vacuum energy gets reduced to a size of the order of $\Lambda_{\rm{QCD}}^4$, which then gets reduced to its current size during the QCD PT. Depending on the UV completion of the SM there may be another GUT and/or SUSY PT (or something else).  A sketch of the evolution of the pressure due to radiation together with that of the vacuum energy (assuming a GUT, EW and QCD PT) is shown in Fig.~\ref{fig:evolution}, which illustrates the main features: 
vacuum energy was much larger at earlier times, nevertheless it always remained a sub-dominant component of the total energy density except around the times of the PTs.  This picture again underlines the interpretation of the cosmological constant as a quantity determined by microscopic physics, as the resulting final vacuum energy   that has changed during the PTs. From the point of view of the cosmological constant problem, this issue is summarized by the equation
\begin{equation}
\Lambda_\text{eff} = \Lambda_\text{bare} +\sum_i \alpha_i T_{c,i}^4 \ ,
\end{equation}
where $\Lambda_\text{eff}$ is the currently observed effective cosmological constant of order $(10^{-3}\ \rm{eV})^4$, the $T_{c,i}$ are the various critical temperatures for every PT the Universe went through, the corresponding $\alpha_i$ being determined by the dynamics of the individual PTs, and $\Lambda_\text{bare}$ is the bare cosmological constant that is used to tune the whole sum to its current value. We can see that the tuning of $\Lambda_\text{bare}$ involves tuning against a sum with several contributions of widely different magnitudes, and the final cosmological constant is extremely sensitive to each one of them. Thus while one gets away with tuning a single parameter, this single tuning encodes sensitivities to a large number of independent dynamical parameters. This is what is reflected in Fig. 1 and is necessary for a viable cosmic history of vacuum energy. 

Whether this is indeed the correct picture of the evolution of vacuum energy is one of the most important fundamental questions of physics that is yet to be verified experimentally.\footnote{A potential alternative history (corresponding to that of an adjustment mechanism) would have a vacuum energy that is always very small, except for some spikes during the PTs, though there is no known, successful, implementation of such a mechanism. Other adjustment mechanisms would go as far as invoking non-local and acausal dynamics, see e.g. \cite{ArkaniHamed:2002fu,Kaloper:2014dqa}.} Any such experimental test would also verify the microscopic origin of the cosmological constant as the gravitational effect of the vacuum energy of the quantum field theory of our Universe, and would thus yield a test of the Equivalence Principle for vacuum energy. The difficulty in verifying this picture experimentally is clear: until very recently, vacuum energy was always a sub-leading component of the energy density, and thus was never the main driver of the expansion. Moreover, the most recent known PT is that of QCD, at a temperature $T_c^{\rm{QCD}}\sim 200$ MeV. While this is a relatively low particle physics scale, most of the phenomena  relevant to experimental cosmology (nucleosynthesis, structure formation, CMBR) are sensitive only to temperatures well below the QCD scale. Thus one would need to consider new observables that are potentially sensitive to the details of the QCD or the EW PTs. This is further complicated by the fact that both of these PTs are thought to be quite weak: the QCD PT is a cross-over, while the EW PT in the SM with a 125 GeV Higgs boson is second order. The imprints of such PTs are weaker than those of strongly first order PTs would be. For example a strongly first order PT is expected to lead to the production of gravitational waves (GWs), whose spectrum could potentially be sensitive to the evolution of vacuum energy during the PT \cite{LisaGeraldine}. Since neither of the PTs is expected to be first order, no significant GWs would have been produced.  

In order to experimentally test properties of vacuum energy, we must find systems where vacuum energy contributes a sizable fraction of the total energy. This can be either in a compact system that can be observed today, or at some earlier epoch in the cosmic expansion in the Universe. We will suggest examples of both types in this paper: we will study the effects of vacuum energy inside the core region of neutron stars, and consider the epochs around cosmic PTs, where vacuum energy approaches the energy stored in matter.\footnote{One more interesting system could be that of a superconducting PT where Cooper pairs condense and generate a tiny photon mass. However, the relevant energy scale is set by the mass gap $\Delta \sim k T_c \sim 10^{-4}$ eV, which is much smaller than the actual mass of the superconductor, and thus any effect will be extremely tiny.} The interesting possibility of searching for effects of vacuum energy on the dark matter relic abundance was considered in~\cite{ChungLongWang}.

The argument leading us to consider neutron stars is the following.  Since it is quite difficult to test the evolution of the true vacuum energy of the Universe, one can look for perturbations where the structure of the vacuum is significantly rearranged, yielding a potentially sizable {\em local} shift in vacuum energy. This could happen in the presence of large local densities, when the large density leads to a change in the structure of the VEVs of the fields. A typical example of this sort of PT is thought to be QCD at high densities. As the chemical potential is increased, QCD is expected to go through a series of PTs even at zero temperature: at very high densities a color-flavor locked (CFL) phase should appear, while at intermediate values a non-CFL quark matter phase should be present \cite{Alford:2007xm}. Both of these phases have VEVs different from the ordinary hadronic phase, and therefore one expects the vacuum energy to also be modified. Of course in this case the change in vacuum energy is tied to the presence of a large density (and its accompanying pressure following an equation of state determined by the QCD dynamics), and experimentally the shift of the vacuum energy in the region of large density manifests itself in a change of the equation of state for the matter in the unusual phase of QCD (which we will just call the condensate or condensed phase). Nevertheless this change in the equation of state of the condensate should have observable experimental consequences. Consider for example a neutron star, one of the densest systems in the Universe. 
Given that their central density is expected to go well beyond the nuclear saturation density, it is thought likely to have an exotic quark condensate of this sort at its core. If vacuum energy indeed contributes an additional piece to the pressure of the condensate, then the structure of the whole neutron star will change compared to the situation where no such additional pressure term is present (for example due to a local adjustment mechanism of the vacuum energy). Thus one will obtain differing structures for neutron stars depending on whether a shift in the vacuum energy is locally cancelled or not. A careful measurement of the mass-radius relation $M(R)$ of the neutron star could potentially distinguish between these scenarios, especially if the equation of state for the condensate is eventually precisely determined by QCD simulations.    

In the second part of the paper we consider the epochs around cosmic PTs, when for a short period vacuum energy becomes sizable compared to radiation. This could modify the propagation of primordial GWs, and leave an imprint on its energy spectrum. The well-studied effect of PTs on GWs is to yield a step in the spectrum which is determined  by the number of relativistic degrees of freedom in thermal equilibrium. Vacuum energy can add a peak in the spectrum, if its magnitude becomes comparable to that of radiation. However, such a peak might be washed out if the step due to the change in the number of degrees of freedom is large, which is indeed what is expected to happen for the QCD PT, while for the EW PT the vacuum energy never becomes large enough to produce a peak. However other PTs can potentially produce a peak, and we will show the conditions needed for that to happen. We also consider the possibility that the time scale for a hypothetical adjustment mechanism for the vacuum energy is somewhat longer than that of the PT. In this case vacuum energy will dominate the total energy for a short period after the PT, and will result in a suppression of the modes that entered before the PT started, yielding a much larger step in the energy spectrum of the GWs than in the standard scenario.

Before we discuss the details of our analysis, we want to comment on what exactly we mean by a changing vacuum energy. There are many different types of PTs in nature. Most of them involve a transition between two phases of matter, without actually changing the VEV of the underlying fields. One example is the recombination of electrons and protons into hydrogen atoms, which happens at around $z\sim 1100$ in the evolution of the Universe, and can be thought of as a transition of ordinary matter from a plasma to a gaseous phase. In this process there is a binding energy of 13.6 eV per hydrogen atom, which will appear as a decrease in the energy density of ordinary matter. Nevertheless we would not consider this a change in vacuum energy. The binding energy is localized around the actual H-atoms, and would dilute like ordinary matter in an expanding Universe, while vacuum energy does not actually get diluted. The type of transition we are after is when the VEVs of fields actually change in a region of space by a significant amount,\footnote{Tiny shifts in VEVs are expected due to matter effects and changing binding energies all the time, but these secondary effects are very small.} leading to a change in the vacuum energy. 

Finally, while we will investigate the effect of a hypothetical adjustment mechanism that cancels the vacuum energy associated with the PT, we will not deal with the details of the adjustment mechanism: we simply assume that it cancels the vacuum energy. Of course one can imagine other potential adjustment mechanisms, which will require modifications of the analysis presented here.  

The paper is organized as follows. In Section~\ref{sec:ns} we present our analysis of the effect of vacuum energy on the structure of neutron stars.  Section~\ref{sec:GW} contains the discussion of the consequences of vacuum energy on the primordial gravitational wave spectrum. We first present some of the general properties of the propagation of gravitational waves in Sec.~\ref{sec:GWgeneral} (while some more related details are in Appendix~\ref{app:gwenergy}). The description of the effects of phase transitions is contained in Sec.~\ref{sec:PT}, while the numerical results for the QCD phase transitions are in Sec.~\ref{sec:QCDandEW}. In Sec.~\ref{sec:PQ} we present the conditions and an example for the case when a peak appears in the gravitational wave spectrum, while the discussion of the effects of an adjustment mechanism can be found in Sec.~\ref{sec:QCDwithadjustment}. Finally we conclude in Section~\ref{sec:Conclusions}. 

\section{Vacuum energy and the structure of neutron stars\label{sec:ns}}
\setcounter{equation}{0}
\setcounter{footnote}{0}

In this section we present our analysis of the effects of vacuum energy on the structure of neutron stars. We will present a toy model for a neutron star, with just two regions: the inner core corresponding to the high-density QCD condensate phase, where the vacuum energy is different from that of low-temperature and low-density QCD, and an outer core in the conventional hadronic phase, with the same condensates that appear all through space since the temperature of the Universe dropped bellow about $T_c^{\rm{QCD}} \sim 200 \MeV$.  This outer region of the star is usually treated as a fluid made of neutrons (and protons and electrons), with a polytropic equation of state (EoS) with no extra vacuum energy. Realistic neutron star simulations are of course much more involved, with many more layers matched onto each other. We are essentially neglecting the crust, the envelope and the atmosphere of the neutron star. We are not attempting to present a precise description of a neutron star, rather to establish the importance of the QCD-scale vacuum energy at the center in contrast to the outer regions. We will show that it has a significant effect on the structure of the star, which would change significantly if the jump in vacuum energy in the inner core was actually not present.  See Ref.~\cite{Lattimer:2012nd} for a review of the physics of neutron stars.

We are assuming a static neutron star in equilibrium at close to zero temperature. Gravitational pressure is balanced by the degeneracy pressure of the fluid. The general form of the metric of a static and spherically symmetric spacetime is given by
\begin{equation}
ds^2=e^{\nu(r)}dt^2- \left(1-2 G m(r)/r\right)^{-1}dr^2-r^2d\Omega^2\ .
\end{equation} 
Einstein's equations for a static and spherically symmetric configuration of a fluid with pressure $p(r)$ and energy density $\rho(r)$ are given by the  Tolman-Oppenheimer-Volkoff equations \cite{Oppenheimer:1939ne,Weinberg:gravitycosmo}:
\begin{eqnarray}
&m^\prime (r)= &4\pi r^2 \rho(r) \,, \label{meq} \\ 
& p^\prime(r)=  &-\frac{p(r)+\rho(r)}{r\left(r-2G m(r) \right)}\left[Gm(r)+4\pi r^3 p(r)\right]\,, \label{peq} \\
&\nu^\prime(r)= &-\frac{2 p^\prime(r)}{p(r)+\rho(r)}\,, \label{nueq}
\end{eqnarray} 
where $^\prime$ denotes differentiation with respect to the radial coordinate $r$. These are three equations for four unknown functions: $p(r)$, $\rho(r)$, $m(r)$ and $\nu(r)$. The extra equation needed to solve the system is the EoS, $p=p(\rho)$, which is the only model dependent input sensitive to the actual phase of the fluid in the various layers of the neutron star. The radius  of the neutron star, $R$, is determined by the condition of vanishing pressure $p(R)=0$. 
Outside the radius of the neutron star, $r> R$, the solution is matched to the Schwarzschild solution in radial coordinates, with total mass $M = m(R)$. 

We model the fluid and its corresponding EoS in the following way: as the pressure increases toward the center of the neutron star, it eventually reaches a critical value $p_{\rm{cr}}$, at some critical surface $r=r_{\rm{cr}}$, where the fluid undergoes a phase transition, from a hadronic phase to a quark matter phase, the latter with a non-vanishing vacuum energy $\Lambda$. To the critical pressure corresponds a density above nuclear saturation, $\sim (200 \MeV)^4$, where nucleons seize to be a good description. There are therefore two EoS's for the two different regions:
\begin{align}
& p = p_{-}(\rho_{-}) \, ,\quad p  \geq p_{\rm{cr}} \, , \quad  r  \leq r_{\rm{cr}} \\
& p = p_{+}(\rho_{+}) \, ,\quad p \leq p_{\rm{cr}} \, , \quad  r  \geq r_{\rm{cr}} \, .
\end{align}
The usual Israel junction conditions \cite{Israel} of continuity of the induced metric and extrinsic curvature at the critical surface require $\nu^\prime(r)$ and $m(r)$ to be continuous across the phase transition. These in turn imply the continuity of the pressure $p(r)$.\footnote{We are neglecting a possible localized surface tension on the layer separating the two phases,  which would allow for a small discontinuity in the pressure at the critical surface.} The energy density $\rho$ is in general discontinuous at $r_{\rm{cr}}$ as is generically the case for phases separated by a spacelike surface, such as the vapor-liquid phases of water.

In the inner core region $r<r_{\rm{cr}}$ we take a polytropic fluid supplemented by a non-vanishing vacuum energy $\Lambda$
\begin{equation}
p_{-} = p_{m} - \Lambda \, , \quad \rho_{-} = \rho_m + \Lambda \, , \qquad p_{m} =  \kappa_{-} \rho_{m}^{\gamma_{-}} \, ,
\end{equation}
where $\rho_{m}$ and $p_{m}$ represent the ordinary matter partial density and pressure, that could include e.g.~the effect of binding energy and interactions, but not the vacuum energy.
In the outer core region, $r>r_{\rm{cr}}$, we take another polytropic fluid described by $\kappa_{+}$ and $\gamma_{+}$ but no vacuum energy, $\Lambda_{+}=0$, that is simply
\begin{equation}
p_{+} = \kappa_{+} \rho_{+}^{\gamma_{+}} \, .
\end{equation}
We will restrict our attention to an outer polytropic EoS with mean exponent $\gamma_+ = 5/3$, which reproduces the low pressure and density limit of a degenerate Fermi gas, and fix the compressibility factor $\kappa_+$ to match nuclear saturation pressure and density, $p_s = (65 \MeV)^4$ and $\rho_s = (185 \MeV)^4$ respectively. For the inner polytropic, we will assume $\gamma_- = 1$ and $\kappa_- = 0.1$ as an approximate description of relativistic quark matter (this is an EoS close to the MIT bag model).\\

\begin{figure}[!t]
\begin{center}
\includegraphics[width=3in]{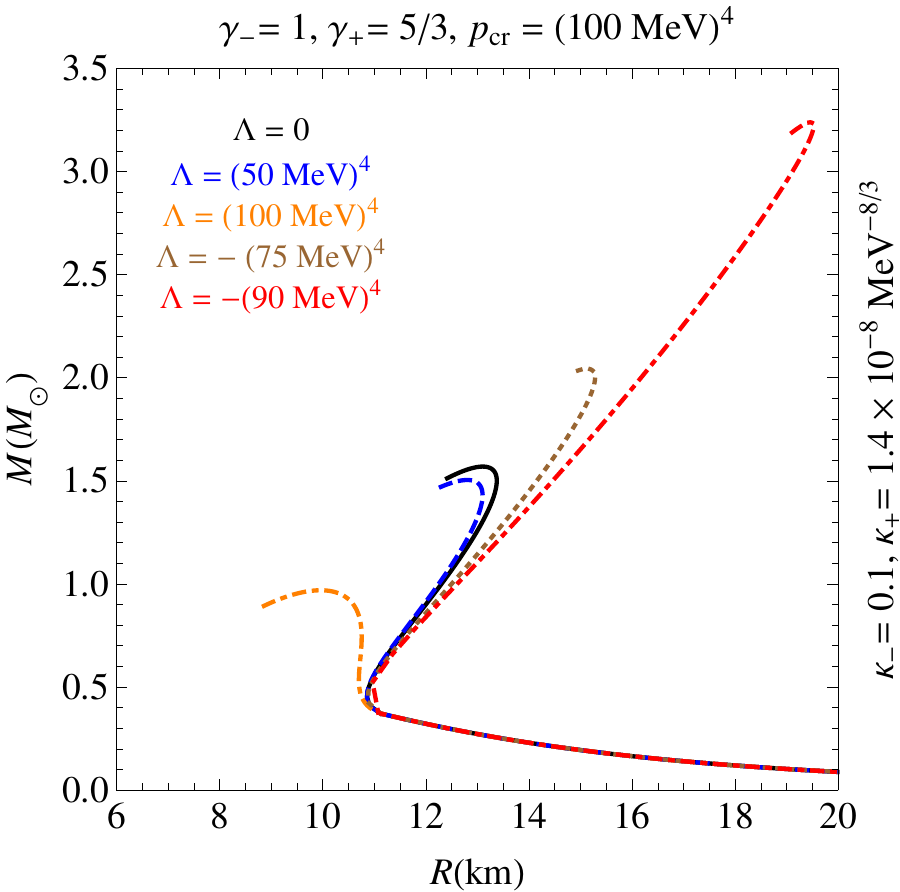}
\caption{Radius versus mass trajectories for a two-polytropic fluid with $\gamma_{-}=1$ and $\gamma_{+}=5/3$, for various values of the vacuum energy in the inner core, $\mathrm{sign}(\Lambda)|\Lambda|^{1/4}=-90 \MeV$ (red dot-dashed), $-75 \MeV$ (brown dotted), $0 \MeV$ (black solid), $50 \MeV$ (blue dashed), and $100 \MeV$ (orange dot-dashed).}
\label{fig:RMplot}
\end{center}
\end{figure}

In addition, we will impose some restrictions on the vacuum energy of the inner phase. On the one hand, should $\Lambda$ be smaller than $-p_{\rm{cr}}$, the matter partial pressure $p_{m}$ would become negative, triggering an instability of the fluid that would split in more than two phases of matter. Thus one has the condition 
\begin{equation}
\label{instab1}
\Lambda> -p_{\rm{cr}} \, .
\end{equation}
On the other hand, we will require the equilibrium configurations obtained after solving Eqs.~(\ref{meq}), (\ref{peq}) and (\ref{nueq}) to be stable. The transition from stability to instability as we vary the pressure at the center of the star, $p_0 = p(r = 0)$, takes place when $\partial M/\partial p_0 = 0$. For the EoS's at hand, it can be shown that we can avoid a stationary point for the total mass of the star if at the transition between the inner and outer fluid the energy density jump is positive,
\begin{equation}
\label{instab2}
\rho_{+} (r_{\rm{cr}}) - \rho_{-}(r_{\rm{cr}}) = \left(\frac{p_{\rm{cr}}}{\kappa_+}\right)^{1/\gamma_+} -\left[\left(\frac{p_{\rm{cr}}+\Lambda}{\kappa_-}\right)^{1/\gamma_-} + \Lambda \right] \geq 0\ .
\end{equation}
This condition imposes an upper bound on the value of the vacuum energy, which depends on $p_{\rm{cr}}$ and the EoS's parameters.

\begin{figure}[!t]
\begin{center}
\includegraphics[width=3in]{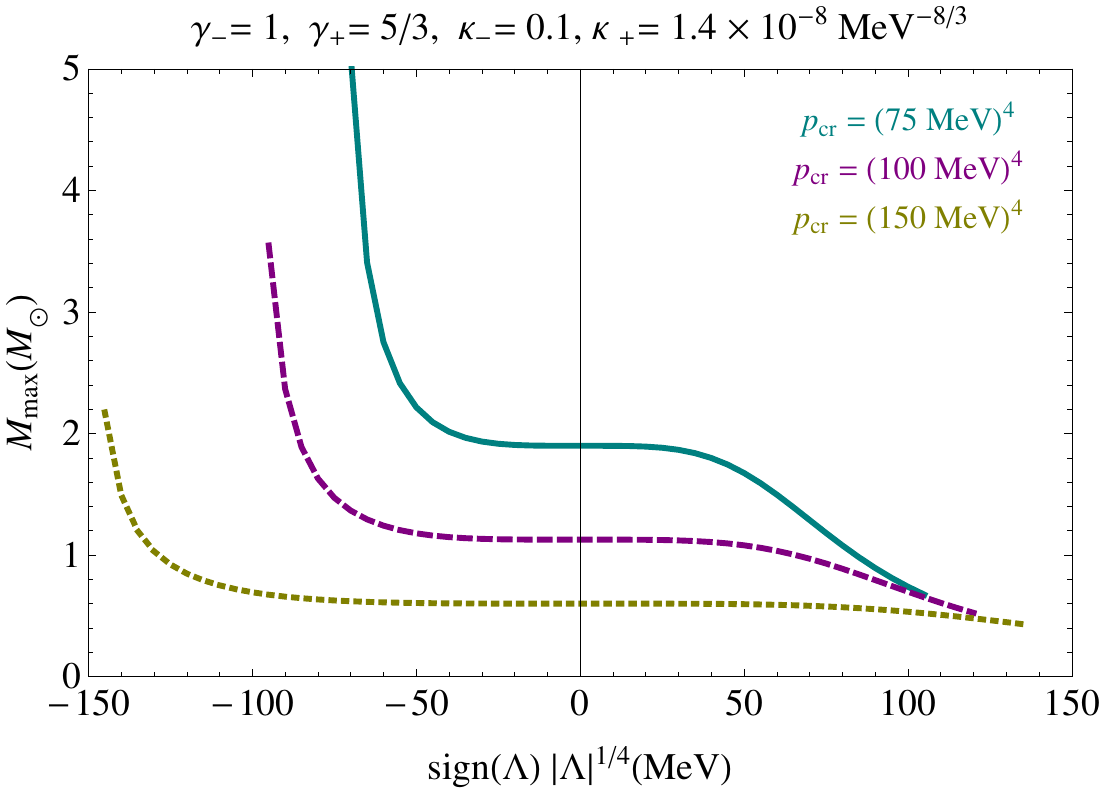}
\includegraphics[width=3in]{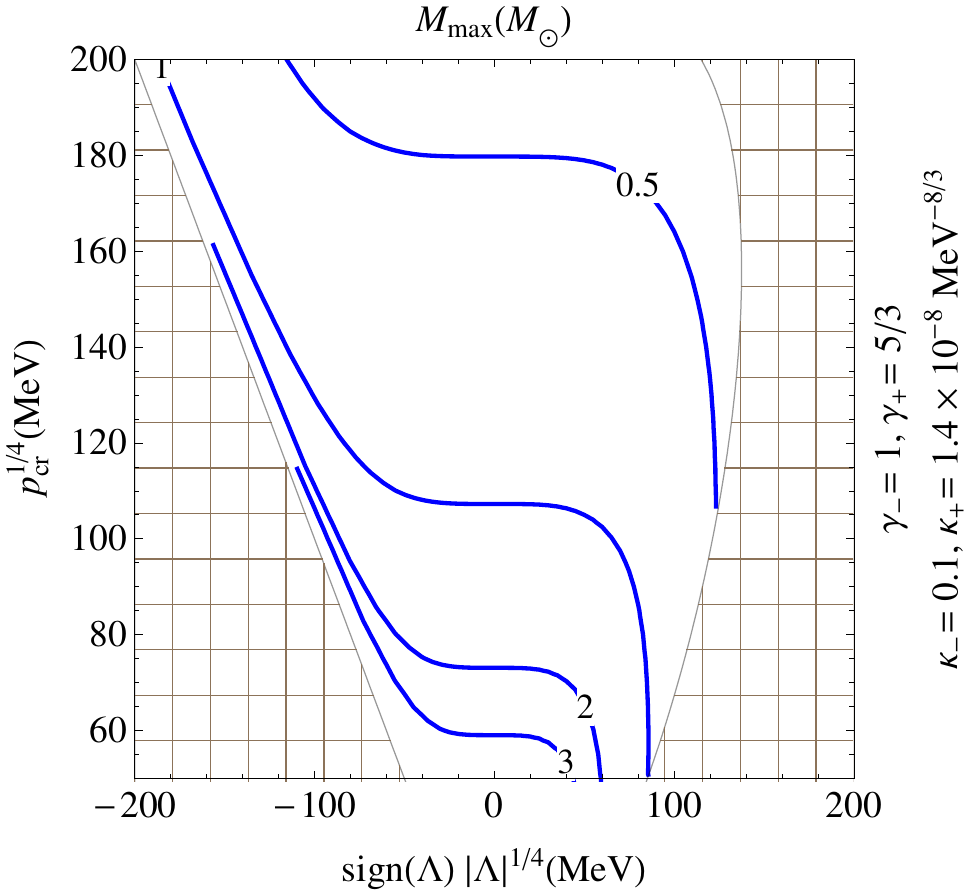}
\caption{Left: Maximum mass with varying $\Lambda$ for several values of the critical pressure $p_{\rm{cr}}$. Right: Maximum mass contour lines in the vacuum energy and critical pressure plane. The shaded regions corresponds to the instability conditions set by  Eqs.~(\ref{instab1}) and (\ref{instab2}). }
\label{fig:Mmax}
\end{center}
\end{figure}
\begin{figure}[!t]
\begin{center}
\includegraphics[width=3.5in]{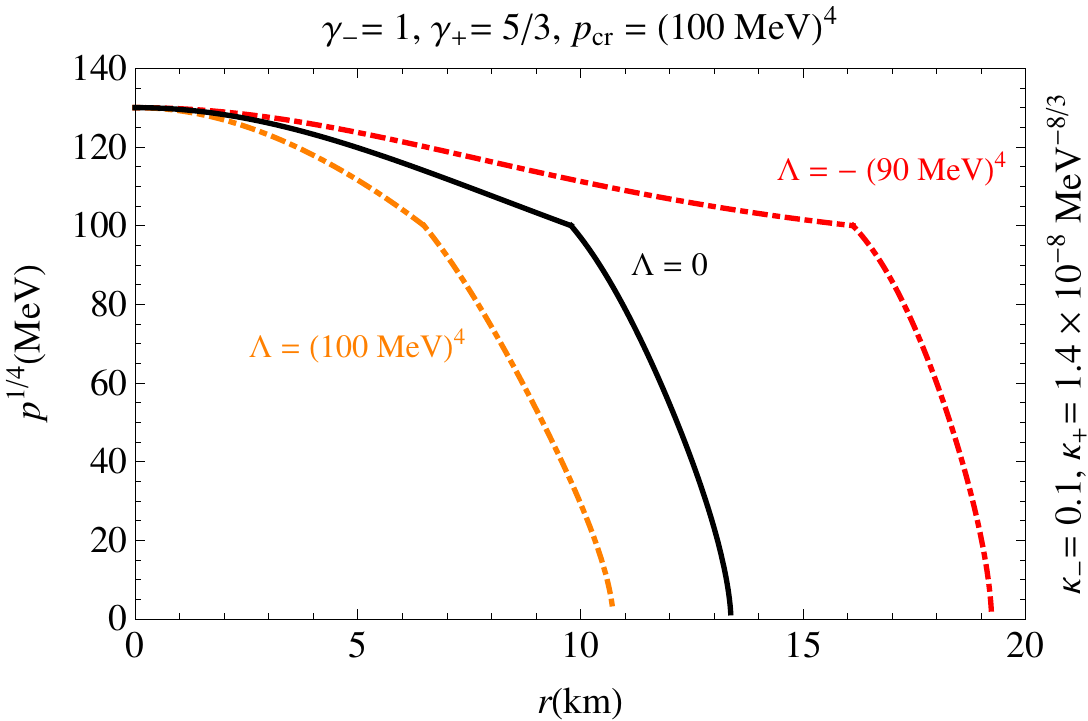}
\caption{Pressure profiles of three stars with the same properties except the value of the inner core vacuum energy, $\mathrm{sign}(\Lambda)|\Lambda|^{1/4}=-90 \MeV$ (red dot-dashed), $0 \MeV$ (black solid), and $100 \MeV$ (orange dot-dashed).}
\label{fig:pprofile}
\end{center}
\end{figure}

In Fig.~\ref{fig:RMplot} we show a representative set of radius versus mass curves for different values of the vacuum energy. We have taken a critical pressure above nuclear saturation, $p_{cr} = (100 \MeV)^4$. Each trajectory has been obtained by varying the central pressure, $p_0$. As the central pressure increases, so does the mass of the star, until it reaches its maximum.
Notice that all the curves converge at low masses, since the outer EoS does not depend on $\Lambda$. Most importantly, note that there is a significant variation of the maximum mass depending on the value of $\Lambda$.
This fact can be best appreciated in Fig.~\ref{fig:Mmax}. 
We show in the left panel several curves of maximum mass as a function of $\Lambda$ for several values of the critical density $p_{\rm{cr}}^{1/4} = 75, 100, 150 \MeV$, while the right panel shows contours of the maximal mass as a function of the vacuum energy and the critical pressure.
If $\Lambda$ is of the order of $p_{\rm{cr}}$, then there is a sizable difference between the maximal possible neutron star mass in the presence of $\Lambda$, as expected in the standard picture, versus $\Lambda \sim 0$ as one would expect for a case with a local adjustment mechanism for the vacuum energy. Furthermore, the sensitivity is higher for smaller values of the critical pressure, as well as for negative values of $\Lambda$. 
Depending on the parameters chosen, an up to 50\% effect can be observed. However, there is generically a long plateau around $\Lambda =0$, implying that for low values of $\Lambda$ when compared to $p_{\rm{cr}}$, the effect of turning off the vacuum energy is small.
Finally, the behavior observed in the figures, in particular the reduction of the maximal mass with $\Lambda$ for a given critical pressure, can be understood by noticing that a larger value of $\Lambda$ implies a larger matter pressure for the same total pressure at the center. This makes the star end at a smaller radius, and hence it has a lower mass. We show in Fig.~\ref{fig:pprofile} the pressure profile of two stars with the same properties except the value of $\Lambda$ to illustrate this point.\\

At this point it is important to take into consideration the fact that there is strong observational evidence of neutron stars with masses above $2M_{\odot}$. Such large masses have been taken as an indication in favor of pure hadronic neutron stars, given the difficulty of reproducing them with EoS's like the MIT bag model.\footnote{See however Ref.~\cite{Fraga:2013qra} for a more refined EoS for quark matter including interactions, and from which higher maximal masses can be obtained.}
We are showing here that if the vacuum energy, which is presumably included in the MIT bag model as part of the bag constant, was to be relaxed towards negligible values, larger values of $M_{\rm{max}}$ could easily be obtained, improving consistency with observations.
Nevertheless, it is certainly crucial that a reliable EoS for the matter component is obtained, before making any definitive conclusions.

With the expected improvement in quantity and quality of experimental data on neutron stars, one might hope to obtain better lower bounds on the maximum mass of a neutron star, along with crucial information on the associated radius. To date, radius measurements have low accuracy, and they have only been achieved for a handful of neutron stars in binary systems, and inferred from X-ray measurements. A promising avenue that is expected to provide new data is the detection of gravitational waves from in-spiraling binary neutron stars \cite{Lackey:2014fwa}.  From the properties of the gravitational wave signatures during coalescence, different competing models for the EoS of the neutron star can be distinguished from one another.  Properties such as the mass-radius relationship, and the response of the star to tidal forces are imprinted on the ``chirp" gravitational wave signature given off by the collapsing binary pair.  Given input from theoretical studies of QCD at high densities where the non-CFL phase is expected to occur, and of the nuclear superfluid equation of state that describes the physics of the outer core, Advanced LIGO could test whether or not there are non-standard contributions to the EoS that may be related to dynamics responsible for the small observed value of the vacuum energy density.  The most challenging aspect of this program, however, is to obtain this theoretical input.  Progress on first-principles determination of the finite chemical potential portion of the QCD phase diagram has been slow, as the typical tools for non-perturbative studies, i.e. the lattice, are ill-suited for large baryon densities.  Further development of experimental techniques to determine properties of exotic phases of QCD, along with the aforementioned advances in theoretical predictions are key to determining the gravitational properties of vacuum energy in neutron stars.

\section{Vacuum energy and primordial gravitational waves\label{sec:GW}}
\setcounter{equation}{0}
\setcounter{footnote}{0}

In this section we investigate the effects of vacuum energy on the propagation of primordial gravitational waves (GWs). Since vacuum energy is comparatively sizable only around the cosmic phase transitions (PTs), those epochs will be the focus of our attention. A main goal of future GW detector experiments (either space based \cite{Cornish:2002fg} or using atom interferometry \cite{Graham:2012sy}) should be to explore the frequency regimes corresponding to the QCD and EW PTs, as well as to look for signals of other possible PTs and search for potential effects of vacuum energy as described below.

\subsection{General properties of the gravitational wave spectrum\label{sec:GWgeneral}}

GWs correspond to transverse traceless tensor perturbations $h_{ij}$ (with  $h_i^i=0$, and  $\partial_k h^k_i=0$) of the metric in an expanding Universe
\begin{equation}
ds^2=a(\tau)^2\left(d\tau^2-(\delta_{ij}+h_{ij})dx^i dx^j \right) \, ,
\end{equation}
where we have used conformal time $\tau$, related to co-moving time $t$ via $a(\tau)d\tau=dt$. The expansion equations in conformal time are given by 
\begin{equation}
\label{confFried}
{a^\prime}^2=(a \dot{a})^2=(a^2 H)^2 = \frac{8 \pi G}{3} a^4 \rho \,,\qquad \frac{a^{\prime\prime}}{a}=a^2\left(\frac{\ddot{a}}{a}+\frac{\dot{a}^2}{a^2}\right)=\frac{4\pi G}{3}a^2T_\mu^\mu\,.
\end{equation}
where $H=\dot{a}/a$ is the Hubble scale with respect to time $t$, $^\prime$ indicates a derivative in $\tau$, and the trace of the energy momentum tensor is $T_\mu^\mu = \rho - 3 p$.
The linearized Einstein equation for the tensor perturbations $h_{ij}$ (assuming no anisotropic stress in the perturbed $T_{\mu\nu}$) is
\begin{equation}
h^{\prime\prime}_{ij}+ 2\mathcal{H} h^{\prime}_{ij} -\nabla^2h_{ij}=0
\end{equation}
where $\mathcal{H}=a^\prime/a$ is the Hubble parameter with respect to conformal time $\tau$. The spatial Fourier transform provides the mode expansion of the gravitational waves:
\begin{equation}
h_{ij}=\sum_{\sigma=+,-} \int \frac{d^3k}{(2\pi)^3} \epsilon^{(\sigma)}_{ij} h^{(\sigma)}_k(\tau) e^{ikx}
\end{equation}
and the evolution equation for the rescaled modes  (omitting the polarization index $\sigma$) 
$\chi_k\equiv a h_k$ becomes
\begin{equation}
\chi^{\prime\prime}_k+ \left( k^2-\frac{a^{\prime\prime}}{a} \right)\chi_k=\chi^{\prime\prime}_k+\left[k^2-\frac{4\pi G}{3}a^2T^\mu_\mu\right]\chi_k=0\,.
\label{eq:basicgw}
\end{equation}
where in the second expression we used Eq.~(\ref{confFried}). Thus the crucial quantity which determines the detailed properties of the GW spectrum is the trace of the energy momentum tensor. 

The basic properties~\cite{Turner} of the solution to Eq.~(\ref{eq:basicgw}) can be understood quite easily. As long as the $k^2$ term dominates the damping term, $\chi$ will freely oscillate, and hence the full solution $h = \chi /a$ will be damped by the scale factor. This damping starts when the given mode enters the horizon.\footnote{It is the real Hubble horizon that a mode $k$ has to enter for the damping to start, $k \gtrsim a H$. When the solution enters the ``horizon'' calculated from $T^\mu_\mu$ in Eq.~(\ref{eq:basicgw}), $k > a \sqrt{4\pi G T^\mu_\mu/3}$, it still has a large velocity, which will start decreasing only when the actual Hubble horizon is entered, in accordance with our expectations from causality.} Before that the mode is frozen, which corresponds to the solution $\chi (\tau ) \propto a(\tau )$ of the equation $\chi''/\chi = a''/a$. Thus the spectrum will be determined by the rate of entering the Hubble horizon. The details of the definitions and evolutions~\cite{Komatsu} of the relevant quantities characterizing GWs are discussed in App.~\ref{app:gwenergy}, where we explain that the energy density per log scale in units of the critical density is approximately given by 
\beq
\Omega_h(\tau, k) \simeq \frac{(\Delta_{h}^{P})^2}{12 H^2(\tau) a^4(\tau)} k^2 a^2(\tau_{hc}) \,,
\label{eq:spectrum}
\eeq
 where $\tau_{hc}$ is the time of horizon crossing, and $\Delta_{h}^{P}$ is the (approximately constant) primordial power spectrum. We can see that the relevant quantity is $k^2 a^2 (\tau_{hc})$.

\subsection{Effects of a phase transition\label{sec:PT}}

If we now consider a mode that enters during radiation domination, when $a \propto \tau$, $H \propto a^{-2}$, then we find that the energy spectrum $\Omega_h$ is flat:%
\footnote{Here we are neglecting the trace anomaly, which makes the equation of state of radiation deviate from the pure conformal behavior $p/\rho = 1/3$, and whose effect on the GW spectrum is to introduce a slight tilt~\cite{Boyle:2005se}.}
\beq
\Omega_h (k>k_{eq}) \propto k^2 a^2(\tau_{hc}) \propto a^4(\tau_{hc}) H_{hc}^2 \propto {\rm const}.
\eeq
since the condition of re-entry is $k^2 \simeq a^2H^2|_{\tau_{hc}}$. 
Here and in the following we will drop the overall factors in Eq.~(\ref{eq:spectrum}) that are common for all the modes.
Also, by $k>k_{eq}$ we mean modes that enter before matter-radiation equality.\footnote{Modes entering during the matter dominated era have a spectrum that scales as $1/k^2$.}

If however, there is a PT then there is a departure from pure radiation domination, and one expects features to show up in the spectrum. The traditional discussion of second order PTs assumes thermal equilibrium and conservation of entropy, with a changing number of relativistic degrees of freedom in thermal equilibrium $g_*(T)$.\footnote{Whether entropy is conserved in a PT where vacuum energy is reduced depends in large part on how quickly the PT proceeds. A nice analogy is to consider a bath with expanding walls, where a compressed spring is also inserted between the walls, where the spring plays the role of the vacuum energy. If the walls expand very quickly while releasing the spring, the spring will start oscillating and its energy eventually is dissipated into the bath. In this case entropy increases, the energy of the spring will go directly into heating the bath, a case analogous to reheating at the end of inflation. However, if the walls expand very slowly, then the spring will slowly relax to its equilibrium position without oscillations and decouple from the system. In this case entropy is conserved and the process is reversible. This is the analog of the scenario usually considered for the QCD and EW PTs.} In this case entropy conservation implies
\begin{equation}
S=\frac{\rho +p}{T}a^3 = {\rm const}. 
\end{equation}
while the number of degrees of freedom determine $\rho + p \propto g_*(T)T^4$. Thus the expansion rate is set by 
\begin{equation}
\label{aAndTentropy}
a \propto T^{-1} g_*^{-\frac{1}{3}}\ .
\end{equation}
This will set the GW spectrum to be 
\beq
\Omega_h (k>k_{eq}) \propto k^2 a^2(\tau_{hc}) \propto a^4(\tau_{hc}) H_{hc}^2\propto g_*^{-\frac{1}{3}}
\eeq
dependent only on the number of degrees of freedom $g_*$. Therefore one expects to see a step in the GW energy spectrum during a PT, of size of approximately $(g_*^b/g_*^a)^{\frac{1}{3}}$  \cite{Schwarz:1997gv}, where $a$ and $b$ denote after or before the PT.

This analysis of PTs so far ignores the potential effects of vacuum energy. Next we will discuss qualitatively what those could look like, while later on we will present the full numerical results for the case of the QCD PT. 

Let us define $\xi(\tau)$ as the relative size of the vacuum energy $\rho_\Lambda$ compared to radiation $\rho_R=\bar{\rho}_R a^{-4}$:
\begin{equation}
\xi = \frac{\rho_\Lambda}{\rho_R}=\frac{\rho_\Lambda}{\bar{\rho}_R}a^4(\tau)\,.
\end{equation}
$\bar{\rho}_R$ carries the dependence on the degrees of freedom $g_*$. Both radiation and vacuum energy set the comoving horizon, which determines the re-entry of the mode $k$, 
\beq
k^2=a^2 H^2 =(1+\xi) a^2\rho_R=(1+\xi)a^{-2}\bar{\rho}_R\,,
\label{eq:horizon}
\eeq
in units where $8\pi G/3 =1$.
The resulting power spectrum is thus 
\begin{equation}
\Omega(k> k_{eq})\propto a^2(\tau_{hc}) k^2=(1+\xi)\bar{\rho}_R=(1+\xi)g_* T^4 a^4\propto (1+\xi)g^{-\frac{1}{3}}_* \,,
\label{eq:generalcase}
\end{equation}
where in the last step we used  entropy conservation, Eq.~(\ref{aAndTentropy}). This is the equation that controls the non-trivial features of the GWs spectrum generated by adiabatic PTs 
where generically both $\xi$ and $g_*$ change, affecting the otherwise flat (or standard) spectrum. Since well before the PT starts $\xi$ is very small and after the PT $\xi$ has to be small again, while during the PT $\xi$ will become sizable, one expects that the effect of the vacuum energy on its own is to produce a peak in the spectrum. Whether this peak will remain as an observable feature will depend on the relative magnitude of the peak (controlled by $\xi$) versus the size of the step (controlled by the change in the number of degrees of freedom). Below we present a discussion of the approximate shape of the expected peak. Those only interested in the actual shape of the spectrum for the QCD PT or for a hypothetical $SU(N)/SU(N-1)$ PT, may skip ahead to Sec.~\ref{sec:QCDandEW} or Sec.~\ref{sec:PQ} respectively.\\

The general expression of the energy spectrum based on Eq.~(\ref{eq:generalcase}) is given by 
\begin{equation}
\label{scalingOmega}
\frac{\Omega(k_a> k_{eq})}{\Omega(k_b> k_{eq})}=\frac{1+\xi_a}{1+\xi_b}\left(\frac{g^a_*}{g^b_*}\right)^{-\frac{1}{3}} \,.
\end{equation}
The generic label in  $\xi_a=\xi(\tau_a)$ and $g_*^a=g_*(\tau_a)$ refers to the mode $k_a=a(\tau_a)H(\tau_a)$ that crosses the horizon at $\tau=\tau_a$. In the following we call $\tau_t$ the starting time of the PT,  $k_t$ the mode entering at that moment, and  $\xi_t$ the vacuum-to-radiation  energy ratio $\xi(\tau_t)$.

Well before and after the PT, where $\xi\rightarrow 0$, we recover the standard flat spectrum with an overall step between the asymptotic values of magnitude $(g^a_*/g^b_*)^{-\frac{1}{3}}$. However, around the PT the vacuum energy is non-negligible and so is $\xi$. The frequency dependence carried by $\xi$ can be exposed inverting
\begin{equation}
k^2 = \sqrt{\rho_\Lambda \bar{\rho}_R}\left(\frac{\xi+1}{\sqrt{\xi}}\right) \simeq \sqrt{\frac{\rho_\Lambda \bar{\rho}_R}{\xi}} 
\end{equation} 
where the last step holds only for $\xi\ll 1$. From this expression and Eq.~(\ref{scalingOmega}) we  see that before the PT, that is before the number of degrees of freedom changes, $g_*= {\rm const}.$, the spectrum scales as 
\beq
\frac{\Omega_h (k \gtrsim k_{t})}{\Omega_h (k \gg k_{t})} \propto \left[ 1+\xi_t \left(\frac{k_t}{k}\right)^4 \right] \,,
\label{eq:klargerkt1}
\eeq
for $\xi_t\ll 1$, implying an increasing spectrum as $k$ approaches $k_t$ from larger values. 

In order to illustrate the qualitative effect on the spectrum of the vacuum energy for $k\lesssim k_t$, let us consider the case with no change in the number of degrees of freedom during the PT, $g^b_* = g^a_*$. Assuming for simplicity that the PT proceeds very quickly, the vacuum energy jumps at $\tau=\tau_t$ from its initial value $\rho_\Lambda$ to zero, and the horizon jumps, consistently with entropy conservation, from $k_t^2$ to $k_t^{\prime\,2}=k^2_t/(1+\xi_t)$. The resulting scale factor $a(\tau_{hc})$ is approximately $k$-independent for the modes $k_t^\prime<k<k_t$, which enter the horizon all at once. Therefore, the spectrum  $\Omega_h \propto k^2 a^2 (\tau_{hc})$, decreases as $k^{2}$ for $k\lesssim k_t$,\footnote{For a slower PT $a(\tau_{hc})$ is a monotonic function of $k$, thus $\Omega_h(k\lesssim k_\tau)$ is expected to decrease slower than $k^2$, the quadratic behavior reached only for an instantaneous PT.}
\beq
\frac{\Omega_h (k_t^\prime <k\lesssim k_t)}{\Omega_h (k=k_t)} =\left(\frac{k}{k_t}\right)^2(1+\xi_t)\,.
\label{eq:kdownkt}
\eeq
One thus expects a peak of size of approximately  $1+\xi_t$. Whether this peak is indeed visible in the spectrum will then depend on the change in the number of degrees of freedom, which gives rise to a step. If $\xi_t \ll (g_{*}^b/g_{*}^a)^\frac{1}{3}-1$ then the peak will be washed out by the step in the spectrum due to the large change in the number of degrees of freedom. However if the change in degrees of freedom is modest such that $\xi_t \gg (g_{*}^b/g_{*}^a)^\frac{1}{3}-1$ then a genuine peak is indeed expected. 
However, if entropy is to be conserved while the number of degrees of freedom change during 
a PT, the Universe must expand by the factor $a(\tau_a)/a(\tau_b) = (T_b/T_a) (g_*^b/g_*^a)^{\frac{1}{3}}$, where $T_b$ is the temperature at the start of the PT, and $T_a$ at the end. This implies that the drop of the spectrum for the modes with $k \lesssim k_t$ will be slower than that in Eq.~(\ref{eq:kdownkt}).

Below in Sec.~\ref{sec:QCDandEW} we will show that for the standard QCD and EW PTs the peak is indeed washed out and one only expects a step. In Sec.~\ref{sec:PQ} we will show an example of a hypothetical PT where the effect of vacuum energy is to produce a peak in the GW spectrum. Finally we will show the case of the QCD PT with an adjustment mechanism whose time scale is longer than that of the PT, resulting in a short period of late inflation in Sec.~\ref{sec:QCDwithadjustment}.

\subsection{Effects of vacuum energy during the QCD and EW phase transitions\label{sec:QCDandEW}}

 Above we have presented the general qualitative picture of the effects of PTs, and in particular the effect of vacuum energy, on the GW spectrum. Here we show the results of the numerical simulation for the QCD PT, and comment on the EW PT as well.  For the case of the QCD PT one can use the results of lattice simulations to learn about the details of the PT, and in particular to read off the effect of vacuum energy. A simple parametrization that has been used in~\cite{latticeQCDPT} is for the trace of the energy momentum tensor
\begin{equation}
\Theta = T^\mu_\mu =  T^4 \left( 1-\frac{1}{(1+e^{(T-c_1)/c_2})^2} \right) \left( \frac{d_2}{T^2}+\frac{d_4}{T^4} \right)
\label{eq:traceT}
\end{equation}
where this parametrization is applicable for 100 MeV $<T<$ 1 GeV, and the approximate values of the constants are $c_1= 193 \MeV$, $c_2=13.6 \MeV$, $d_2 = 0.241\GeV^2$, $d_4= 0.0035 \GeV^4$. The meaning of these parameters is quite clear: $c_1$ is the critical temperature, $c_2$ is the characteristic temperature width of the PT, $d_4$ corresponds to the QCD vacuum energy, while $d_2$ is a matter density present during the PT. Eq.~(\ref{eq:traceT}) determines the equation of state of the QCD matter during the PT, modified from pure radiation. Since $\Theta = \rho -3p$, using the first law of thermodynamics $\rho = T (dp/dT) - p$, we can obtain the pressure as
\begin{equation}
p(T) = p_0 \frac{T^4}{T_0^4} + T^4 \int_{T_0}^T dT' \frac{\Theta (T')}{T'^5}\ .
\end{equation}
One can choose $T_0=1$ GeV, and assume that at those temperatures the pressure arises from pure QCD radiation $p_0= \frac{\pi^2}{90} g_* T_0^4$, with $g_* \approx 55$. 
The energy density is then obtained from the knowledge of $\Theta = \rho -3p$ and $p$.

\begin{figure}[!t]
\begin{center}
\includegraphics[width=4in]{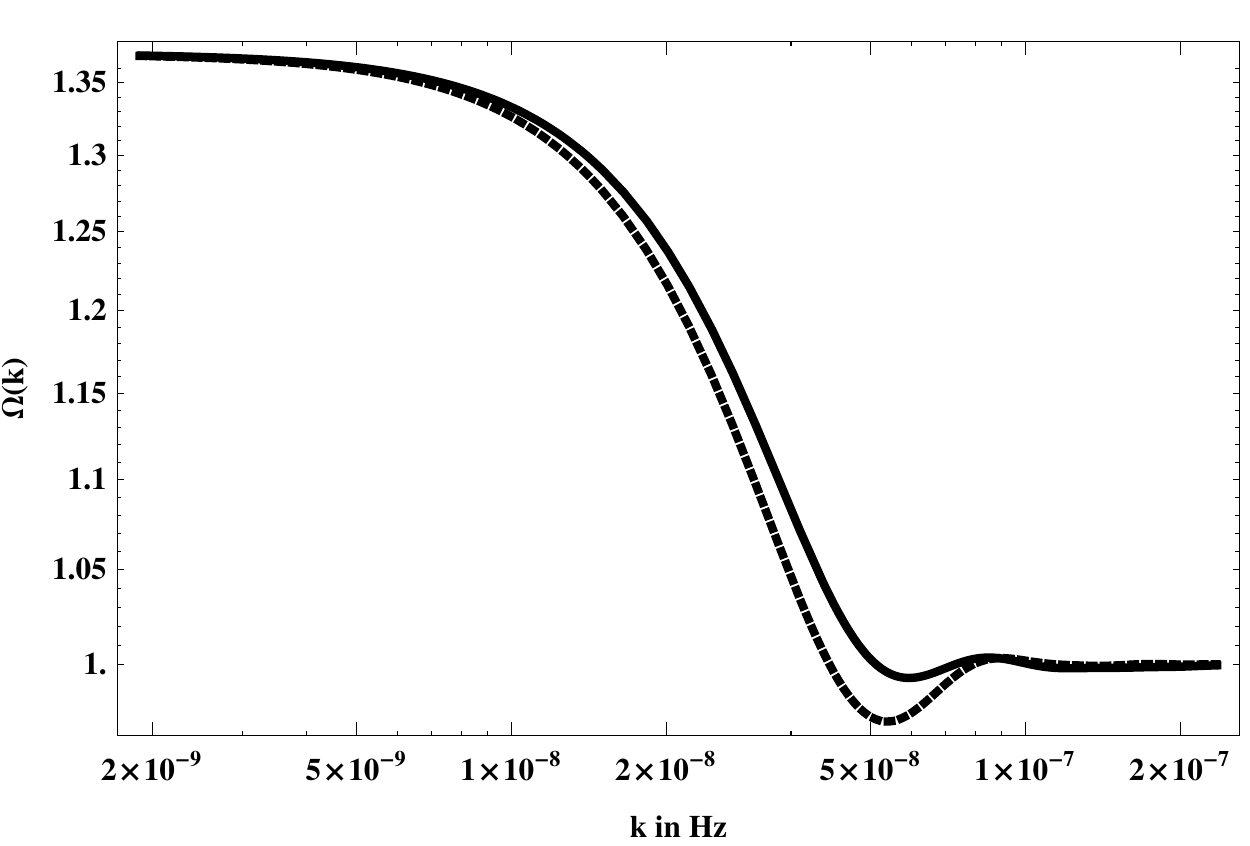}
\caption{The effect of the QCD PT on a flat primordial GW spectrum from a numerical simulation. The solid curve uses the results of the lattice simulations Eq.~(\ref{eq:traceT}), which includes the effect of vacuum energy. The dashed curve corresponds to a modified equation of state describing the QCD PT, where the coefficient $d_4$ describing vacuum energy has been increased (while $d_2$ set to zero) such that the number of degrees of freedom remains the same after the end of the PT. We can see that neither curve contains a sharp peak, but rather the expected step in the spectrum due to the change in the number of degrees of freedom. There is only a small change in the detailed shape of the spectrum, while another curve with no vacuum energy and only $d_2$ turned on would lie right on top of the solid curve.}
\label{fig:QCDPT}
\end{center}
\end{figure}

Assuming that the PT proceeds sufficiently slowly, and that entropy is conserved during the QCD PT,\footnote{This is not a trivial assumption. We will see later in Sec.~\ref{sec:PQ} that if the change in the vacuum energy is too large during a PT (with the change in the number of degrees of freedom fixed), then entropy can not be conserved. We have checked that with the given equations of state the QCD PT does proceed sufficiently slowly such that it might be adiabatic.} we can determine the scale factor $a(T)$ as a function of the temperature:
\begin{equation}
a(T) = \left( \frac{T}{T_0} \right)^\frac{1}{3} \left( \frac{\rho (T_0)+p(T_0)}{\rho (T)+p(T)} \right)^\frac{1}{3} a(T_0) \ .
\end{equation}
Finally, the temperature as a function of the conformal time $\tau$ is obtained from integrating the Friedman equation:
\begin{equation}
\tau (T) = \tau_0 +\int_{T_0}^{T} dT' \frac{da(T')}{dT^\prime}\frac{1}{a^2(T')} \frac{1}{\sqrt{8\pi G \rho (T')/3}} \ .
\end{equation}
This latest integral can be performed numerically, leading to a numerical function of $T(\tau )$, which in turn can be used to determine the input function $4\pi G T^\mu_\mu/3$ in Eq.~(\ref{eq:basicgw}). This can then be used to numerically study the spectrum of GWs over the QCD PT as follows. We assume that a particular mode prior to entering the horizon was just given by a plain sine function (the solution to the free equation), so for the boundary condition of the numerical solution to the differential equation we will use $\chi_k (\tau_0)= \sin (k \tau_0)/k$ and $\chi'_k(\tau_0) = \cos (k \tau_0)$, with arbitrary overall normalization, however the $k$ dependent factor is included in order to reproduce a flat primordial spectrum. Once the PT is over, we match the $\chi$ function again to sines and cosines $\chi_k(\tau ) = (A_k \sin (k\tau )+B_k\cos (k\tau ))/k$. The energy spectrum will then be given by $\Omega_h (k) = (A_k^2 +B_k^2)/k^2$. The results of the simulation are given in Fig.~\ref{fig:QCDPT}. We can see that there is no peak appearing in the spectrum: the step due to the change in the number of degrees of freedom during the QCD PT from approximately 51.25 down to 17.25 completely covers up the small effect of the vacuum energy. This in in accordance with our qualitative expectations from the previous section. There we argued that the magnitude of the peak is set by $\xi_t = \rho_\Lambda / \rho_R(\tau_t)$. Here we can identify $\rho_\Lambda = d_4/4$, and for $\rho_{R}(\tau_t)$ we take the value of radiation at the PT temperature $\rho_{R}(\tau_t) \approx 0.025 \GeV^4$, leading to an estimated peak size $\xi_t \approx 0.04$. On the other hand the magnitude of the step for QCD is given by $(g_*^b/g_*^a)^\frac{1}{3} -1 \approx 0.43$. The large step, of order 43\%, covers up the peak of the order of a few percent. In fact we have tried to see how robust this answer is to the details of the QCD PT, by modifying the relative magnitudes of $d_2$ and $d_4$. One extreme case would be when $d_2 =0$, and $d_4$ is chosen such that the number of degrees of freedom still matches the QCD value at the end of the PT. We can see in Fig.~\ref{fig:QCDPT} that increasing the value of $d_4$ to the  maximal possible value does not change the basic features of the GW spectrum: there are small distortions in the details, but the basic shape dominated by the large step remains unchanged, and no peak appears in either case. 

Similarly, no peak is expected for the case of the EW PT in the Standard Model. The reason for this is again the very large number of degrees of freedom, coupled with the fact that the Higgs mass is quite low, giving rise to a small vacuum energy. The vacuum energy before the PT is 
\begin{equation}
\rho_\Lambda = \frac{1}{8} m_h^2 v^2 \simeq (105\,{\rm GeV})^4
\end{equation}
because $\langle |H|^2\rangle_{T>T_c}=0$.
The  critical temperature $T_c\simeq m_h v/m_{top}\simeq 175 \GeV$, where the coefficient of the $|H|^2$ term in the Higgs potential vanishes, is mainly determined by the top quark's thermal loop contribution $\delta m_H^2(T)=y_t^2 T^2/4$. With $g_*=106.75$ before the PT, we find 
\begin{equation}
\rho_R(T_c)=\frac{\pi^2}{30}g_* T_c^4 \simeq  (425\, {\rm GeV})^4\,,
\end{equation}
resulting in $\xi_t \simeq 0.004$, a tiny peak compared to the expected step of order 7\% due to the change in the number of degrees of freedom from $106.75$ to $86.25$.

\subsection{Conditions for a peak in the spectrum\label{sec:PQ}}

We have seen above that the effect of vacuum energy on the GW spectrum is quite small during the QCD and EW PTs. The reasons for this can be summarized as follows. There are a large number of degrees of freedom, which will make the relative contribution of vacuum energy small, if the couplings are perturbative. For non-perturbative couplings like for the case of QCD, one still needs to make sure that the change in the number of degrees of freedom does not overwhelm the effect of vacuum energy. 

We can look for conditions on the details of a PT such that a peak actually remains visible in the GW spectrum. Since the total entropy is proportional to $\rho +p$, and the first law of thermodynamics tells us that this is equal to  $T \, dp/dT$, we require
$dp/dT >0$.
This condition is equivalent to 
\begin{equation}
\label{temporary1}
-\frac{1}{p_R}\frac{dp_\Lambda}{dT}< \left(\frac{1}{g_*}\frac{dg_*}{dT}+\frac{4}{T}\right)
\end{equation}
where we parametrized the pressure  as $p_R=\pi^2 g_*(T)T^4/90$ for radiation in terms of the number of effective degrees of freedom $g_*(T)$ at temperature $T$. At linear order in temperature change $\Delta T$, and recalling that $\Delta p_\Lambda=-\Delta\rho_\Lambda= -\rho_\Lambda$ and $p_R= \rho_R / 3$ at the beginning and at the end of the PT, we get\footnote{Here we have neglected  the contribution from changing the equation of state parameter $w$. Including such contributions, the condition (\ref{temporary1}) would be modified to
\begin{equation}
\label{temporary2}
-w_\Lambda \frac{1}{\rho_R}\frac{d\rho_\Lambda}{dT}\left(1+\frac{d\log w_\Lambda}{d\log T}/\frac{d\log \rho_\Lambda}{d\log T}\right)< w_R \left(\frac{1}{g_*}\frac{dg_*}{dT}+\frac{4}{T}\right)\,. \nonumber
\end{equation}
However, we expect the extra term proportional to $dw_{R, \Lambda}/dT \neq 0$ to be actually harmless. Considering e.g. the limit where $dw_\Lambda/dT$ becomes the dominant term in the bracket, the condition becomes 
\begin{equation}
\left(-\frac{\Delta w_\Lambda}{\Delta T} \right)\frac{\rho_\Lambda}{\rho_R}< w_R \left(\frac{1}{g_*}\frac{\Delta g_*}{\Delta T}+\frac{4\Delta T}{T}\right)\,.\nonumber
\end{equation}
This is a trivial condition since $\omega_\Lambda=-1$ is the lower bound for sensible EoS's, requiring therefore $\Delta w_{\Lambda}/\Delta T>0$.}
\begin{equation}
\label{condition1}
\frac{\rho_\Lambda}{\rho_R}< \frac{1}{3} \left(\frac{\Delta g_*}{g_*}+\frac{4\Delta T}{T}\right)\,.
\end{equation}
This is an upper bound on $\xi_t$ given the change in the number of degrees of freedom. 
If $ \Delta g_*/g_* \gg 4 \Delta T/T$ then the peak is overwhelmed by the step due to the change in the degrees of freedom, since in this case 
\begin{equation}
\xi_t < \frac{1}{3}\frac{\Delta g_*}{g_*}\simeq \left(g_*^b/g_*^a\right)^\frac{1}{3}-1\,.
\end{equation}
In order to show a visible peak one needs the opposite limit, $ \Delta g_*/g_* \ll 4 \Delta T/T$, that is the fractional change in the number of degrees of freedom is small compared to the relative width of the PT. In this case the positive entropy condition just requires 
\begin{equation}
\xi_t < \frac{4}{3} \frac{\Delta T}{T} \gg \frac{1}{3} \frac{\Delta g_*}{g_*}
\end{equation}
 but a peak can still dominate over the step in the GW spectrum because the step is much smaller than the upper bound on $\xi_t$ set by the fractional change in the temperature.

An extra condition, $d^2p/dT^2>0$, ensures a decreasing temperature in the expanding Universe. However, using $dp/dT=s$, it implies $d\rho/dT>0$ which in turn gives a condition that it is trivially satisfied at the linear level when Eq.~(\ref{condition1}) holds.

\begin{figure}[tb]
\begin{center}
\includegraphics[width=4in]{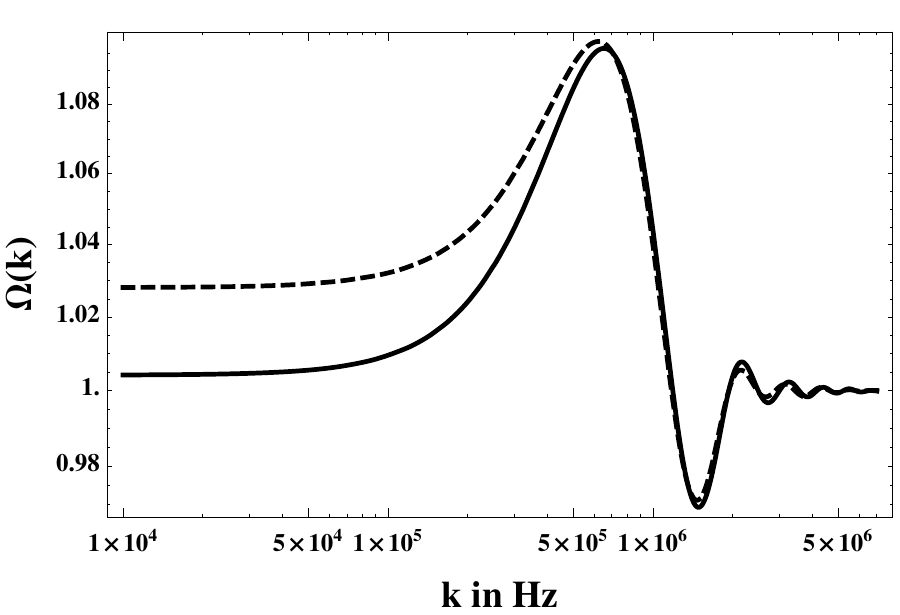}
\caption{The effect of a hypothetical $SU(N)/SU(N-1)$ PT on a flat primordial GW spectrum from a numerical simulation. We set the critical temperature at $T_c=10^{11} \GeV$, a small change in the number of degrees of freedom, giving rise to a 3\% step (dashed), or absent altogether (solid), and a large ratio of vacuum energy to critical temperature, due to a sizable quartic $\lambda \approx 18$ in Eq.~(\ref{xiPQPT}). The peak due to the vacuum energy is clearly observable, both with and without a step. }
\label{fig:PQPT}
\end{center}
\end{figure}

To verify that the peak in the primordial GW spectrum can indeed dominate the step from the change in degrees of freedom, we consider a hypothetical PT corresponding to a high-scale $SU(N)$ symmetry breaking via a complex scalar multiplet $\Phi$. In order to maximize $\xi_t$ compared to any possible step we need to make sure that the change in the number of degrees of freedom is small, and the actual vacuum energy is maximized. Therefore we consider a theory with $N$ complex scalars $\Phi = (\Phi_1, \dots , \Phi_N)$ and a potential with a sizable quartic self-interaction $\lambda$ of the form
\begin{equation}
V(\phi)=\lambda\left(|\Phi|^2-\frac{f^2}{2}\right)^2\ .
\end{equation}
The self-interactions yield a thermal mass contribution $\delta m_{\Phi}^2(T)=\lambda (N+1)T^2/6$ which determines the critical temperature $T_c\simeq 6 f^2/(N+1)$ (and hence the time $\tau_t$) where the PT starts.  The vacuum-to-thermal energy ratio is therefore
\begin{equation}
\xi_t=\frac{\rho_\Lambda(\tau_t)}{\rho_{R}(\tau_t)}=\frac{10}{3}\left(\frac{N+1}{g_*}\right)\times\left(\frac{\lambda (N+1)}{16\pi^2}\right)\,.
\label{xiPQPT}
\end{equation}
The last term in the bracket is bounded by perturbativity to be $O(1)$ or smaller.  Taking $\lambda \approx 18$ and $N=5$, the number of degrees of freedom is $g_*=116.75$, and $\xi_t \approx 0.12$. The magnitude of the step can be quite small, if the masses of the $2N-1$ Goldstone bosons resulting from the breaking of the $SU(N)$ global symmetry to $SU(N-1)$ are much below the critical temperature. In this case the step and the peak can be separated from each other in frequencies and a clean peak is expected to arise. In Fig.~\ref{fig:PQPT} we show an example where the step and the peak are separated, and hence only the peak is visible at frequencies corresponding to the PT, as well as a case where they are both present, but the peak dominates over the step. 

These GW spectra have been obtained by modeling the hypothetical PT, specifically the pressure, with an ansatz compatible with the constraints (\ref{temporary1}) and (\ref{temporary2}). The behavior of $p(T)$ interpolates (by means of $\tanh$ functions) between the asymptotic EoS's before and after the PT, $p(T\gg T_c)=g_{*}^b \pi^2 T^4/90-\lambda f^4/4$ and $p(T\ll T_c)=g_{*}^a \pi^2T^4/90$ respectively. We have extracted the energy density $\rho(T)$ during the phase transition from $dp/dT=(\rho+p)/T$, computed the trace of the energy momentum tensor $T^\mu_\mu = \rho - 3 p$, and then followed the same procedure presented in the previous subsection for the calculation of the spectrum.

Finally, we would like to note that a similar spectrum is obtained when considering a Peccei-Quinn $U(1)$ symmetry breaking via two sets of coupled scalars: a complex scalar $\phi$, whose VEV breaks spontaneously the PQ symmetry and its self-coupling $\lambda$ sets the resulting vacuum energy, and additional $N$ complex scalars $\tilde{\phi}_i , i=1, \ldots ,N$ whose large coupling $\lambda'$ to $\phi$ set the critical temperature. A large ratio $\lambda'/\lambda$ allows relatively large values of $\xi_t$, and a sizable peak in the GW spectrum.

\subsection{Effects of an adjustment mechanism\label{sec:QCDwithadjustment}}

It is conceivable that the dynamics of an adjustment mechanism for the vacuum energy completely changes the character of the cosmological PTs associated with QCD, EW, or other high temperature vacuum re-arrangements.  In particular, the time scale of vacuum energy adjustment may be significant in comparison with the elapsed cosmological time over which these PTs usually take place.  In such cases, after the PT, the vacuum energy associated with the high temperature phase would be temporarily stored in the sector associated with the relaxation of the vacuum energy.  A short period of inflation is then possible after each PT, during which the vacuum energy is slowly released.  A reheating mechanism would also be necessary, with the temperature of reheating being lower than the critical temperature for the PT.

Such inflationary epochs would strongly suppress the amplitude of the GW modes which had already entered the cosmological horizon prior to the PT.  The factor by which they are suppressed is approximately $r \sim (a_0/a_f)^4$, where $a_0$ is the scale factor at the beginning of the inflationary regime, and $a_f$ is the scale factor when it ends.  Modes which are outside the horizon during this short inflationary era are simply frozen, and remain immune to the rapidly growing scale factor~\cite{Schettler:2010dp}.

In order to study the possible effects of an adjustment mechanism, we model a PT (such as that associated with QCD) by assuming a high temperature phase during which the pressure of the fluid is given by pure radiation
\begin{equation}
p_\text{high}(T) = \frac{\pi^2}{90} g_*^\text{high} T^4. 
\label{eq:phigh}
\end{equation}
For the temperature of the PT, we use $T_c = 198 \MeV$.  Ordinarily it is assumed that there is a vacuum energy that drops across the PT, being converted adiabatically into cosmic expansion and/or lower temperature radiation.  If there is an adjustment mechanism at work, however, energy could be transferred to a sector that is not in thermal equilibrium with standard model fields.  In this case, there would be a vacuum energy that carries an explicit time dependence, much like it does in standard early Universe inflationary models:
\begin{equation}
\rho_\Lambda^\text{after} = \rho_\Lambda (t).
\end{equation}
The form of this time dependence is model dependent, but the presumption in this section is that it decreases slowly, and that the time scale for this relaxation is significant compared with the cosmic time over which the PT usually takes place.  For the purpose of this analysis, we take the energy density after the PT to consist of pure vacuum energy that remains constant for some co-moving time $\Delta t = t_\text{relax}$, after which it drops quickly, being replaced by radiation at some low temperature $T= T_\text{reheat} < T_c$.

\begin{figure}[t!]
\begin{center}
\includegraphics[width=4in]{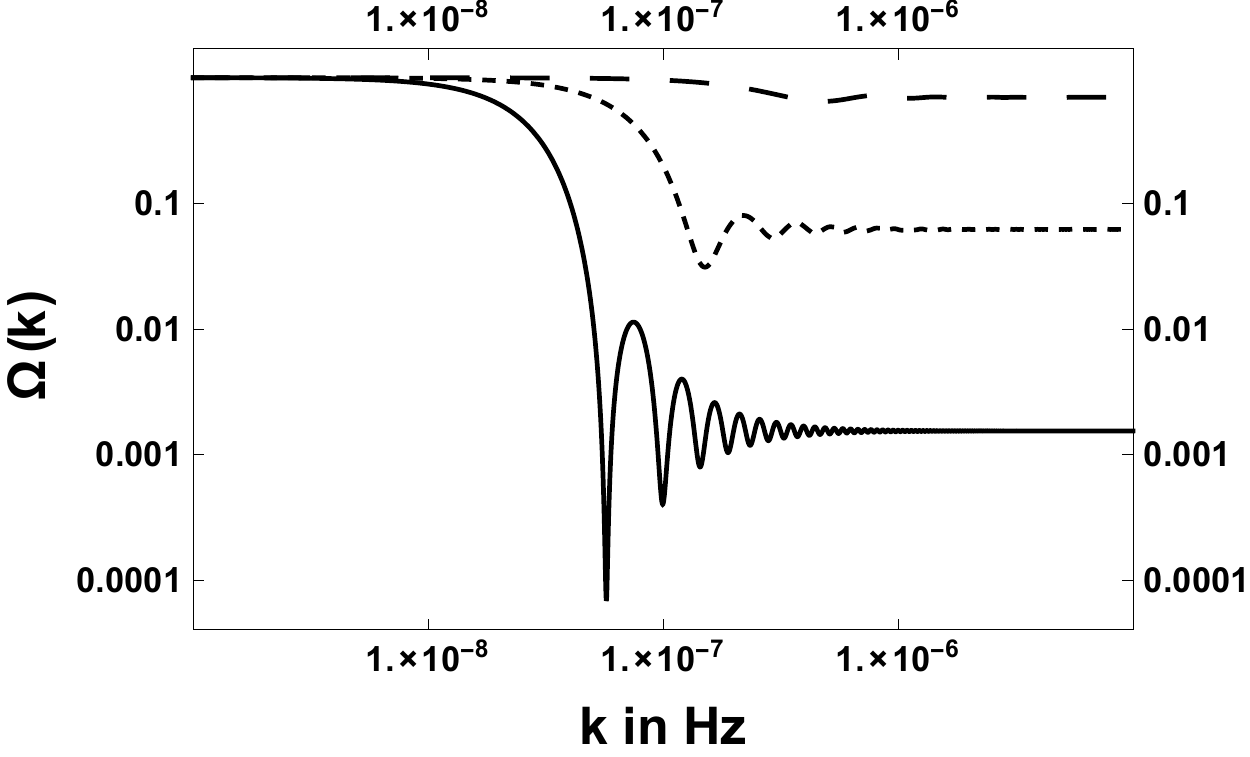}
\caption{The effect of an adjustment mechanism on the spectrum of GWs that are entering the cosmological horizon near the time of the QCD PT.  Curves are normalized to 1 at low wave number.  The three curves correspond to epochs of inflation that last for different amounts of co-moving time given by $t_\text{relax} = t_\text{QCD}$ (long dash), $5 \, t_\text{QCD}$ (short dash), and $10 \, t_\text{QCD}$ (solid), where $t_\text{QCD}$ is the age of the Universe in co-moving time at the QCD PT.  The reheat temperature is taken to be $T_\text{reheat} = T_c/10$. }
\label{fig:QCDinflation}
\end{center}
\end{figure}

In summary, we envision the following alternative history of the QCD PT to be:
\begin{itemize}
\item At $T> T_\text{QCD}$, the system is in thermal equilibrium, with the pressure given by pure radiation, as in Eq.~(\ref{eq:phigh}).  This is the history up until comoving time $t = t_\text{QCD}$.
\item At times $t$ satisfying $t_\text{QCD} < t < t_\text{QCD}+t_\text{relax}$, the Universe is dominated by vacuum energy, which we assume to be constant: $p = -\rho_\Lambda$.  Based on lattice studies of high temperature QCD near the cross-over, we take $\rho_\Lambda^\text{QCD} = (243 \MeV )^4$.  We note that this choice is model dependent.  The details of the spectrum will be sensitive to the way in which the adjustment sector couples to QCD dynamics.
\item At times $t > t_\text{QCD}+ t_\text{relax}$, we presume a reheating has occurred, and the Universe is again radiation dominated and in thermal equilibrium at some temperature $T_\text{reheat}$ below the critical temperature associated with the PT.
\end{itemize} 
We again solve the wave equation governing the evolution of primordial GWs in the early Universe, only now the vacuum energy acts as an explicitly time dependent mass term.  We numerically solve for the evolution for various wave numbers, presuming a scale invariant primordial spectrum, with the results displayed in Fig.~\ref{fig:QCDinflation}.  

The position of the step on the $k$-axis depends on the amount of inflation.  This is because reheating occurs at different values of the scale factor in each scenario.  The physical frequency of the modes that would be observed by GW experiments scales like $1/a_\text{now}$, which varies in each case, shifting the location of the step.  If one were to reduce the reheating temperature further, $a_\text{now}$ would shrink in order to maintain the correct values of the currently observed fluid densities, and the steps would move to higher frequencies.  The reheat temperature however cannot be too low,  as this would interfere with big bang nucleosynthesis.  For the EW transition, the bound on the reheat temperature would be less severe, and there would be more freedom in the position of the step.

The assumption of equilibrium in the previous cases related the size of the step to the change in the number of degrees of freedom:  $\Delta \Omega_h/\Omega_h \approx (g_*^b/g_*^a)^{\frac{1}{3}} -1$, which for QCD is approximately $0.43$.  For the case of out-of-equilibrium dynamics associated with a dynamical adjustment of the vacuum energy, we find that much larger steps are possible.  Observation of anomalously large steps in the GW spectrum at wave numbers associated with cosmological PTs is a possible indicator of a dynamics that may play a role in stabilizing a small value for the vacuum energy.

There might be other interesting types of PTs that are each worthwhile to study in the context of signals of vacuum energy or effects of adjustment mechanisms. One often discussed PT is that of the breaking of the conformal symmetry in RS1 models. The PT itself is expected to be first order, and should produce GWs on its own~\cite{LisaGeraldine}. In addition, the details of this PT might significantly influence the dynamics of the other PTs, such as EW and QCD. The nature of the EW PT might also change significantly if embedded into a larger theory like supersymmetry. All of these are interesting directions that should be explored in connection with possible signals of vacuum energy.

\section{Conclusions\label{sec:Conclusions}}
\setcounter{equation}{0}

Vacuum energy changes during phase transitions, and has settled to its current value only after the QCD phase transition (or later). While at earlier epochs vacuum energy was much larger than its current value, it was always a subleading component of the total energy density, except perhaps around the times of the phase transitions. Confirmation of this picture would provide major support of the multiverse scenario, and rejection of it would call into question the necessity of anthropic arguments for the smallness of the cosmological constant. In this paper we have proposed potential experimental tests for two different aspects of vacuum energy in phases different from ours. The core of neutron stars may contain a non-standard phase of QCD, in which vacuum energy is expected to contribute an $O(1)$ fraction of the total energy. We presented a simple model of neutron stars which demonstrated that vacuum energy can significantly affect their mass-radius relation.  A careful measurement of this, together with a first principles theoretical determination of the equation of state of the various phases of QCD can provide evidence for the presence of vacuum energy at the core of the neutron stars. Our second approach is more directly related to the cosmic evolution of vacuum energy. Here we propose that a careful measurement of the primordial gravitational wave spectrum at frequencies corresponding to phase transitions may contain interesting information about the nature of the changing vacuum energy during the phase transition. While we expect there not to be a signal from the standard model QCD and EW phase transitions, an adjustment mechanism might change this significantly. If the adjustment time scale is much larger than that of the phase transition, there would be a significant suppression of the higher frequencies of the gravitational wave spectrum. We also demonstrated that additional phase transitions might show up as peaks in the spectrum.  This connection of primordial gravitational wave signals with the dynamics of vacuum energy provides a strong additional motivation for planning and building more sensitive gravitational wave experiments testing different frequency bands.   While measuring the effects of vacuum energy is quite challenging, the importance of this issue warrants that all stops be pulled for eventually completing the program of the verification of the cosmic history and gravitational effects of vacuum energy.

\section*{Acknowledgments}
\setcounter{equation}{0}
\setcounter{footnote}{0}

We thank Chiara Caprini, Eanna Flanagan, Anson Hook, Nemanja Kaloper, Andrew Long, Juan Maldacena, Matt Reece, Lorenzo Sorbo, and Filippo Vernizzi for useful discussions. B.B.,  C.C., J.H., and J.T. thank the Aspen Center for Physics for its hospitality while some of this work was performed.   B.B. thanks the Mainz Institute for Theoretical Physics (MITP) for its hospitality during the completion of this work. C.C. thanks the CERN theory group for its hospitality while this work was in progress. J.H. thanks Cornell University for hospitality throughout this work. B.B and J.S. are supported in part by the MIUR-FIRB grant RBFR12H1MW.  B.B. is also supported in part by the Agence Nationale de la Recherche under contract ANR 2010 BLANC 0413 01, and by the ERC Starting Grant agreement 278234 ``NewDark" project. C.C. is supported in part by the NSF grant PHY-1316222.  J.H. is supported by the DOE under grant DE-FG02-85ER40237. J.T. was supported in part by the DOE under grant DE-SC-000999.

\appendix
\section*{Appendix}

\section{Energy density in gravitational waves\label{app:gwenergy}}
\setcounter{equation}{0}
\setcounter{footnote}{0}

We define $h_{\sigma, k}$ as the Fourier transform of the metric perturbation 
\beq
\vev{ h_{\sigma,\mathbf{k}} h_{\sigma',\mathbf{k'}} } = (2 \pi)^3 \delta_{\sigma \sigma'} \delta(\mathbf{k} + \mathbf{k'}) |h_{\sigma,k}|^2 \ , \quad h_{ij}(\tau,\mathbf{x}) = \int \frac{d^3 k}{(2\pi)^3} \epsilon_{ij}^\sigma h_{\sigma,\mathbf{k}}(\tau) e^{i \mathbf{k} \cdot \mathbf{x}}\,.
\eeq
The physically relevant quantity characterizing gravitational waves is the energy density at a given conformal time
\beq
\rho_h(\tau) = \frac{1}{16 \pi G a^2(\tau)} \int \frac{d^3 k}{(2\pi)^3} |h'_{\sigma,k}|^2
\eeq
where the integral runs over comoving wave numbers $k$,  and summation over polarizations $\sigma = +, -$ is understood, while the associated power spectrum is given by
\beq
\Delta_h^2 = \frac{4 k^3}{2 \pi^2} |h_{k}|^2 \ , \quad |h_{k}|^2 = |h_{\sigma, k}|^2 \ .
\eeq
 When considering primordial perturbations created during inflation, it is convenient to define the transfer function $\mathcal{T}(\tau, k)$ such that
\beq
h_{k}(\tau) \equiv h_{k}^{P} \mathcal{T}(\tau,k)
\eeq
where the primordial amplitude from inflation $h_{k}^{P}$ has (approximately) constant power,
\beq
(\Delta_{h}^{P})^2 = \frac{4 k^3}{2 \pi^2} |h_{k}^{P}|^2 \simeq \frac{2}{\pi^2} \frac{H_{\star}^2}{M_P^2}
\label{eq:infpower}
\eeq
which remains constant once the modes exit the horizon during inflation. $H_\star$ is the Hubble constant at horizon exit. We then have
\beq
(\Delta_{h})^2 = (\Delta_{h}^{P})^2 \mathcal{T}^2(\tau,k)
\eeq
We can then write the energy density in terms of the transfer function
\beq
\rho_h(\tau) =\int d \ln k \tilde \rho_h(\tau,k) \,,\qquad  \tilde \rho_h(\tau,k) =\frac{ (\Delta_{h}^{P})^2 \mathcal{T}'^2(\tau,k)}{32 \pi G a^2(\tau)}
\label{eq:gravdensity}
\eeq
It is customary to work instead with the energy density per logarithmic scale normalized to the critical density
\beq
\Omega_h(\tau, k) \equiv \frac{\tilde \rho_h(\tau,k)}{\rho_c(\tau)} 
\eeq
where  $\rho_c = 3 H^2(\tau)/8 \pi G$.
Therefore one has
\beq
\Omega_h(\tau, k) = \frac{(\Delta_{h}^{P})^2}{12} \frac{1}{H^2(\tau)}  \frac{1}{a^2(\tau)} \mathcal{T}'^2(\tau,k)
\eeq
It will be convenient for the arguments below to approximate $\mathcal{T}'$ above assuming that the wave modes are deep inside the horizon $k \tau \gg 1$ (or $k \gg aH$), in which case 
\beq
\mathcal{T}'^2(\tau,k) \simeq k^2 \, \mathcal{T}^2(\tau,k) \ .
\label{Tapprox}
\eeq
Based on our discussion on the freeze-out and reentry of modes we can easily understand the basic properties of  $\Omega_h$. All modes become super horizon, $k \ll a H$, 
during inflation, and once outside the horizon their power spectrum $\Delta_{h}^2$ freezes to the value set by inflation Eq.~(\ref{eq:infpower}), independent of $k$.  This means that once a mode reenters the horizon at $\tau = \tau_{hc}$, it does it asymptotically with the same power, irrespective of when it enters. Thus we will approximate $[\mathcal{T}(\tau_{hc},k)]^2 \simeq 1$.
Since gravitons are already decoupled from the thermal bath from the very start of the expansion,   the evolution of the energy density once inside the horizon should scale with the expansion as radiation $\rho_h(\tau) \sim \tilde \rho_h(\tau,k) \sim a^{-4}(\tau)$. This implies, from Eqs.~(\ref{eq:gravdensity}) and (\ref{Tapprox}), that $[\mathcal{T}(\tau < \tau_{hc},k)]^2 \sim a^{-2}(\tau)$. Taking into account the value of the transfer function at horizon crossing, we find
\beq
\mathcal{T}^2(\tau < \tau_{hc},k) \simeq \frac{a^2(\tau_{hc})}{a^2(\tau)}\ .
\eeq
This way we obtain Eq.~(\ref{eq:spectrum}) in this approximation.



\begin{thebibliography}{99}

\bibitem{accelerating}
 A.~G.~Riess {\it et al.}  [Supernova Search Team Collaboration],
  ``Observational evidence from supernovae for an accelerating universe and a cosmological constant,''
  Astron.\ J.\  {\bf 116}, 1009 (1998)
  \arXivold{astro-ph/9805201};
 S.~Perlmutter {\it et al.}  [Supernova Cosmology Project Collaboration],
  ``Measurements of Omega and Lambda from 42 high redshift supernovae,''
  Astrophys.\ J.\  {\bf 517}, 565 (1999)
  \arXivold{astro-ph/9812133}.

\bibitem{Weinbergnogo}
 S.~Weinberg,
  ``The Cosmological Constant Problem,''
  Rev.\ Mod.\ Phys.\  {\bf 61}, 1 (1989).

\bibitem{dilatoncc}
 R.~Contino, A.~Pomarol and R.~Rattazzi, unpublished;
 B.~Bellazzini, C.~Csaki, J.~Hubisz, J.~Serra and J.~Terning,
  ``A Naturally Light Dilaton and a Small Cosmological Constant,''
  Eur.\ Phys.\ J.\ C {\bf 74}, 2790 (2014)
  \arXiv{1305.3919}{hep-th};
 F.~Coradeschi, P.~Lodone, D.~Pappadopulo, R.~Rattazzi and L.~Vitale,
  ``A naturally light dilaton,''
  JHEP {\bf 1311}, 057 (2013)
  \arXiv{1306.4601}{hep-th}.

\bibitem{Bludman:1977zz}
  S.~A.~Bludman and M.~A.~Ruderman,
  ``Induced Cosmological Constant Expected above the Phase Transition Restoring the Broken Symmetry,''
  Phys.\ Rev.\ Lett.\  {\bf 38} (1977) 255.

\bibitem{ArkaniHamed:2002fu}
N.~Arkani-Hamed, S.~Dimopoulos, G.~Dvali and G.~Gabadadze,
``Nonlocal Modification of Gravity and the Cosmological Constant Problem,''
  \arXivold{hep-th/0209227}.
 
\bibitem{Kaloper:2014dqa}
N.~Kaloper and A.~Padilla,
``Vacuum Energy Sequestering: the Framework and Its Cosmological Consequences,''
Phys.\ Rev.\ D {\bf 90} (2014) 8, 084023
[Addendum-ibid.\ D {\bf 90} (2014) 10, 109901]
\arXiv{1406.0711}{hep-th}.

\bibitem{LisaGeraldine}
 L.~Randall and G.~Servant,
  ``Gravitational waves from warped spacetime,''
  JHEP {\bf 0705}, 054 (2007)
  \arXivold{hep-ph/0607158}.

\bibitem{ChungLongWang}
D.~Chung, A.~Long and L.~T.~Wang,
  ``Probing the Cosmological Constant and Phase Transitions with Dark Matter,''
  Phys.\ Rev.\ D {\bf 84}, 043523 (2011)
   \arXiv{arXiv:1104.5034}{astro-ph.CO}.

\bibitem{Alford:2007xm}
  M.~G.~Alford, A.~Schmitt, K.~Rajagopal and T.~Sch\"afer,
  ``Color superconductivity in dense quark matter,''
  Rev.\ Mod.\ Phys.\  {\bf 80} (2008) 1455
  \arXiv{0709.4635}{hep-ph}.

\bibitem{Lattimer:2012nd}
  J.~M.~Lattimer,
  ``The nuclear equation of state and neutron star masses,''
  Ann.\ Rev.\ Nucl.\ Part.\ Sci.\  {\bf 62} (2012) 485
  \arXiv{1305.3510}{nucl-th}.

\bibitem{Oppenheimer:1939ne}
J.~R.~Oppenheimer and G.~M.~Volkoff,
``On Massive Neutron Cores,''
Phys.\ Rev.\ {\bf 55} (1939) 374; 
R.~C.~Tolman,
``Static Solutions of Einstein's Field Equations for Spheres of Fluid,''
Phys.\ Rev.\ {\bf 55} (1939) 364.

\bibitem{Weinberg:gravitycosmo}
see e.g. S.~Weinberg,
``Gravitation and Cosmology: Principles and Applications of the General Theory of Relativity,''
John Wiley \& Sons (1972), 688 p

\bibitem{Israel}
W.~Israel, 
``Singular hypersurfaces and thin shells in general relativity,"
Nuovo Cim. {\bf B44}, 1 (1966). 

\bibitem{Fraga:2013qra}
  E.~S.~Fraga, A.~Kurkela and A.~Vuorinen,
  ``Interacting quark matter equation of state for compact stars,''
  Astrophys.\ J.\  {\bf 781} (2014) 2,  L25
  \arXiv{1311.5154}{nucl-th}.

\bibitem{Lackey:2014fwa}
  B.~D.~Lackey and L.~Wade,
  ``Reconstructing the neutron-star equation of state with gravitational-wave detectors from a realistic population of inspiralling binary neutron stars,''
  \arXiv{1410.8866}{gr-qc}.

\bibitem{Cornish:2002fg}
  N.~J.~Cornish, D.~N.~Spergel and C.~L.~Bennett,
  ``Journey to the edge of time: The GREAT mission,''
  \arXivold{astro-ph/0202001}.

\bibitem{Graham:2012sy}
  P.~W.~Graham, J.~M.~Hogan, M.~A.~Kasevich and S.~Rajendran,
  ``A New Method for Gravitational Wave Detection with Atomic Sensors,''
  Phys.\ Rev.\ Lett.\  {\bf 110} (2013) 171102
  \arXiv{1206.0818}{quant-ph};
  P.~W.~Graham private communication.
  
\bibitem{Turner}
 M.~S.~Turner, M.~J.~White and J.~E.~Lidsey,
  ``Tensor perturbations in inflationary models as a probe of cosmology,''
  Phys.\ Rev.\ D {\bf 48}, 4613 (1993)
 \arXivold{astro-ph/9306029}.

\bibitem{Komatsu}
 Y.~Watanabe and E.~Komatsu,
  ``Improved Calculation of the Primordial Gravitational Wave Spectrum in the Standard Model,''
  Phys.\ Rev.\ D {\bf 73}, 123515 (2006)
  \arXivold{astro-ph/0604176}.

\bibitem{Boyle:2005se}
  L.~A.~Boyle and P.~J.~Steinhardt,
  ``Probing the early universe with inflationary gravitational waves,''
  Phys.\ Rev.\ D {\bf 77} (2008) 063504
  \arXivold{astro-ph/0512014}.

\bibitem{Schwarz:1997gv}
D.~J.~Schwarz,
``Evolution of Gravitational Waves Through Cosmological Transitions,''
Mod.\ Phys.\ Lett.\ A {\bf 13} (1998) 2771
\arXivold{gr-qc/9709027}.

\bibitem{latticeQCDPT}
  A.~Bazavov, T.~Bhattacharya, M.~Cheng, N.~H.~Christ, C.~DeTar, S.~Ejiri, S.~Gottlieb and R.~Gupta {\it et al.},
  ``Equation of state and QCD transition at finite temperature,''
  Phys.\ Rev.\ D {\bf 80} (2009) 014504
  \arXiv{0903.4379}{hep-lat};
S.~Borsanyi, G.~Endrodi, Z.~Fodor, A.~Jakovac, S.~D.~Katz, S.~Krieg, C.~Ratti and K.~K.~Szabo,
  ``The QCD equation of state with dynamical quarks,''
  JHEP {\bf 1011}, 077 (2010)
  \arXiv{1007.2580}{hep-lat};
  R.~R.~Caldwell and S.~S.~Gubser,
  ``Brief history of curvature,''
  Phys.\ Rev.\ D {\bf 87}, no. 6, 063523 (2013)
  \arXiv{1302.1201}{astro-ph.CO}.

\bibitem{Schettler:2010dp} 
  S.~Schettler, T.~Boeckel and J.~Schaffner-Bielich,
  ``Imprints of the QCD Phase Transition on the Spectrum of Gravitational Waves,''
  Phys.\ Rev.\ D {\bf 83}, 064030 (2011)
  \arXiv{1010.4857}{astro-ph.CO}.

\end{thebibliography}
\end{document}